\documentclass[11pt,a4paper]{JHEP}
\usepackage[T1]{fontenc}		
\usepackage[ansinew]{inputenc}
\usepackage{amsmath}
\usepackage{amssymb}
\usepackage{slashed}
\usepackage{graphicx}
\usepackage{subfigure}
\newcommand{\mn}{{\mu\nu}}

\renewcommand{\a}{\alpha}
\renewcommand{\b}{\beta}
\newcommand{\g}{\gamma}
\renewcommand{\d}{\delta}

\newcommand{\e}{\epsilon}

\renewcommand{\k}{\kappa}
\renewcommand{\l}{\lambda}
\newcommand{\m}{\mu}
\newcommand{\n}{\nu}

\renewcommand{\r}{\rho}
\newcommand{\s}{\sigma}

\renewcommand{\L}{\Lambda}

\newcommand{\be}{\begin{equation}}
\newcommand{\ee}{\end{equation}}
\newcommand{\beqa}{\begin{eqnarray}}
\newcommand{\eeqa}{\end{eqnarray}}

\newcommand{\pd}{\partial}
\renewcommand\m{\mu}
\renewcommand\a{\alpha}

\renewcommand\n{\nu}
\renewcommand\b{\beta}	
\renewcommand\r{\rho}
\newcommand\di{\mathrm{d}}
\renewcommand\s{\sigma}

\title{\sc{Scale-invariant alternatives to general relativity}\date{}}
\author{Diego Blas\footnote{diego.blas@epfl.ch}\;, Mikhail Shaposhnikov\footnote{mikhail.shaposhnikov@epfl.ch}\, 
and Daniel Zenh\"ausern\footnote{daniel.zenhaeusern@epfl.ch} \vspace{.2cm}\\
\normalsize%\llap{$^a$}
 \it  Institut de Th\'eorie des Ph\'enom\`enes Physiques,\\
\it \'Ecole Polytechnique F\'ed\'erale de Lausanne,\\
 \normalsize\it CH-1015, Lausanne, Switzerland\\
 }

 \abstract{We study the general class of gravitational field theories 
  constructed on the basis of scale invariance (and therefore 
  absence of any mass parameters) and invariance under
 transverse diffeomorphisms (TDiff),
 which are the 4-volume
 conserving coordinate transformations. 
 We show that these theories are
 equivalent to a specific type of scalar-tensor theories of gravity
 (invariant under all diffeomorphisms) with a number of  properties, making them
 phenomenologically interesting. They contain, in
 addition to the dimensionless coupling constants of the original theory, an
arbitrary dimensionful parameter  $\Lambda_0$. 
  This parameter is
 associated with an integration constant of the equations of motion,
 similar to the arbitrary cosmological constant appearing in unimodular
 gravity.  We focus on the theories where Newton's constant and the electroweak
 scale emerge from the spontaneous breaking of scale invariance and
 are unrelated to $\Lambda_0$.
 The massless particle spectrum of these theories contains
 the graviton and a new particle -- dilaton. For $\Lambda_0=0$, the massless dilaton
 has only derivative couplings to matter fields and the bounds 
 on the existence of a 5th force are easily satisfied. As for the matter fields, we determine the conditions leading to a renormalizable  low-energy theory.
 If $\L_0\neq 0$, scale invariance is  broken. The arbitrary constant $\L_0$ produces a ``run-away'' potential
 for the dilaton.
  As a consequence, the dilaton can act as a dynamical dark energy component.
 We elucidate the origin of the cosmological constant
 in the class of theories under consideration and formulate the condition
 leading to its absence. If this condition is satisfied,
 dark energy is purely dynamical and associated to the dilaton.}

\begin{document}

%%%%%%%%%%%%%%%%%%
\section{Introduction}
\label{intro}
%%%%%%%%%%%%%%%%%
General Relativity (GR) and the Standard Model of particle physics
(SM) are characterized by two very different energy scales: the Planck
mass $M_P$ (related to Newton's constant as $M_P=(8\pi
G_N)^{-1/2}=2.4\cdot10^{18}\mathrm{GeV}$)  in the case of GR and the
vacuum expectation value of the Higgs field in the SM, $v \simeq 250 \,
\mathrm{GeV}$.\footnote{One can add to these scales a cosmological
constant $\L\sim  10^{-47}\ \mathrm{GeV}^4$ whose attribution to one
or the other sector (GR or SM) is not  understood.} In theories where a scalar field $\phi$ interacts ``non-minimally'' with the scalar curvature $R$ through a term $\xi \phi^2 R$, the Planck mass can be generated dynamically \cite{Minkowski:1977aj,Zee:1978wi}.
In such theories, the Planck scale and the electroweak scale may have a common origin.
A minimal option to realize this idea is to identify the scalar field $\phi$ with the Higgs
field of the SM and choose the constant $\xi\sim M_P/ M_W\sim
{10^{16}}$. In this case, the origin of the Planck scale is related to the electroweak
symmetry breaking and the existence of a very large dimensionless constant \cite{CervantesCota:1995tz,vanderBij:1993hx}.
However, in this theory the Higgs field almost completely decouples from the other
fields of the SM \cite{CervantesCota:1995tz,vanderBij:1993hx}, leading
to contradiction with the precision tests of the electroweak theory.
Moreover, the validity of theories with such large dimensionless
parameters remains unclear. Therefore, adding extra fields to the SM and GR
seems unavoidable for the realization of the ``one-scale'' scenario. The
addition of new fields is further motivated by the fact that it allows to implement
the idea of a ``no-scale'' scenario (see below).

In \cite{Shaposhnikov:2008xb} two of us (M. S. and D. Z.) proposed an
extension of the SM and GR containing an extra real scalar field $\chi$ --
dilaton -- and containing no absolute energy scale. Earlier works
with similar ideas, but different in a number of essential
points, include 
\cite{Fujii:1982ms,Wetterich:1987fk,Wetterich:1987fm}. The
Lagrangian of the model was fixed with the following
principles:
\begin{enumerate}
\item[i)] \label{c1} The action does not contain terms with more than two
derivatives.
\item[ii)] \label{c2} The action is invariant under global scale transformations
\begin{align}\label{strans}
\Phi(x)&\mapsto \lambda^{d_\Phi}\Phi(\lambda x),
\end{align} 
where $\Phi$ stands for the different fields in the action (scalar,
spinor, vector and gravitational), $\lambda$ is an arbitrary real
constant and $d_\Phi$ is the canonical mass dimension of the field
$\Phi$.  The dilatational invariance is not preserved by the standard
regularization schemes (such as dimensional regularization,
Pauli-Villars regularization, cut-off regularization or lattice regularization) used in the
Jordan frame formulation of the theory\footnote{In the present context of scale-invariant theories, we define the Jordan frame as the frame in which the action is invariant under scale transformations of the form \eqref{strans}, and where the metric has zero mass dimension.}.
In fact, all these schemes introduce an explicit parameter with dimension
of mass and hence break scale invariance. This eventually 
translates into the presence of anomalies.
However, for the cases
where scale invariance is spontaneously-broken, one can formulate modified regularization schemes that do not
introduce any intrinsic mass
parameter\footnote{Interestingly, these schemes become ``standard'' in
the Einstein frame formulation of the theory
\cite{Bezrukov:2008ej,Bezrukov:2010jz}. See also the comments after Eq. (\ref{eq:tildepot}).}, i.e. that are anomaly-free. The scale-invariant version
of dimensional regularization is discussed in 
\cite{Englert:1976ep,Shaposhnikov:2008xi}, the field-dependent cut-off
in \cite{Wetterich:1987fm} and lattice regularization in
\cite{Shaposhnikov:2008ar}. The resulting effective, rather than
fundamental, field theories \cite{Shaposhnikov:2009nk} are
scale-invariant to all orders of perturbation theory. The existence of scale-invariant regularization schemes suggests that
exact scale invariance  can still be a legitimate guiding principle
for the construction of new theories.
\item[iii)] \label{c3} The particle physics part of the theory given in the Jordan frame is polynomial in the different fields.
\item[iv)] \label{c4} The Higgs-dilaton potential in the Jordan frame contains a flat
direction and leads to spontaneous breaking of scale invariance.
Any vacuum state on this flat direction gives rise to 
identical physics\footnote{It is important to distinguish this way of
spontaneous  breaking of scale invariance from the approach presented
in  \cite{Jain:2008qv}. The authors of \cite{Jain:2008qv} argued that the mere existence 
of cosmological evolution may be enough to provide an energy scale
from which every other mass can be derived. It is, though, unclear whether 
this proposal can be made phenomenologically acceptable.}. The Planck
scale, particle masses, and quantum dimensional transmutation
parameters like $\Lambda_{QCD}$ are generated dynamically.
\item[v)] \label{c5} The space-time metric obeys the constraint $g=-1$,
where $g=\det{g_\mn}$, corresponding to Unimodular Gravity (UG)
\cite{vanderBij:1981ym,
Wilczek:1983as,Zee:1983jg,Buchmuller:1988wx,Weinberg:1988cp,
Unruh:1988in,Alvarez:2005iy,Henneaux:1989zc}, rather than conventional
GR.
\end{enumerate}

Besides realizing the ``no-scale'' scenario, this proposal solves (in
a technical sense) the problem of stability of the Higgs mass against
radiative corrections, which are kept small due to the exact scale
invariance. For the precise meaning of this statement, see \cite{Shaposhnikov:2008xi}. The requirement iv)  leads to the absence of
a cosmological constant term. The spontaneous breakdown of scale invariance naturally provides a mechanism for inflation. In addition, the almost massless and very weakly coupled dilaton acts as dynamical dark energy.

In the present work, we will consider the proposal 
of \cite{Shaposhnikov:2008xb}, and extend it to a setup in which the additional scalar degree of freedom appears as a part of the metric
field. To introduce the setup, let us first recall that the spectrum of GR consist of
 \emph{just}
 massless spin-two degrees of freedom 
 (which is intimately related to the invariance of the theory under
diffeomorphisms, called Diff invariance henceforth)  \cite{Feynman}. The
 only other possible metric theory of gravity sharing this feature is unimodular gravity (UG)  \cite{vanderBij:1981ym,Alvarez:2006uu}.
Both possibilities are nearly equivalent
(at least
at the classical level): any solution of the UG equations of motion corresponds
to a
solution of the GR equations with a particular value of the
cosmological
constant $\Lambda=\Lambda_0$. The only difference is that the parameter
$\Lambda$ in GR is a fundamental constant while $\L_0$ in UG is an integration constant 
determined by initial conditions.
One can go beyond GR by considering the (perhaps) more physical requirement of having a consistent metric theory \emph{including} spin-two polarization.  It
can be proven  \cite{vanderBij:1981ym,Alvarez:2006uu} that the minimal group of gauge invariance
required to construct such a theory is the subgroup of coordinate transformations
\be
\label{TD}
x^\m\mapsto \tilde x^\m(x),\ \mathrm{with} \ J\equiv \left| \frac{\pd
\tilde x^\m}{\pd x^\n}\right|=1\;,
\ee
which is generated by the subalgebra of \emph{transverse} vectors, 
$$
\delta x^\m=\xi^\m, \quad \pd_\m \xi^\m=0.
$$
We will refer to these transformations as \emph{transverse diffeomorphisms}
(TDiff)\footnote{There  are different names for these transformations
in the literature, including volume or area preserving
diffeomorphisms.  In this work we will use the name TDiff
 as it does not make any reference to the  dimensionality
of  space-time. Besides, we will use the term ``TDiff theories'' for theories invariant
under TDiff.}.
Under TDiff transformations the determinant of the metric $g$ transforms as a scalar. Even more, in TDiff theories $g$ generally corresponds to a propagating scalar degree of freedom, the only exceptions being GR and UG \cite{Alvarez:2006uu}.
As has been argued in the past, a TDiff theory is not 
equivalent to standard scalar-tensor gravity but rather to
UG plus a new scalar field, which is potentially massive \cite{Buchmuller:1988wx,Alvarez:2006uu}. The objective of this work is to study general properties of 
 scale-invariant TDiff theories, and to  show that they are phenomenologically viable. We mainly work with the classical theory.

The paper is organized as follows.
We start in Section \ref{tdit} with a discussion of TDiff gravity theories
and provide a simple proof of their classical equivalence 
to Diff invariant theories of gravity with an arbitrary integration constant $\Lambda_0$.  After discussing the
background solutions, we explicitly show that a scalar degree of freedom 
is present in the gravitational sector. From this analysis, we clarify the role of $\L_0$ as 
an extra initial condition  
 representing a new coupling constant of a peculiar potential for the scalar field.

Next, the attention is turned to scale-invariant
TDiff theories (Section \ref{sitd}).
It is shown that TDiff invariance allows to choose potentials that lead to spontaneous breaking of scale invariance and thereby generate all energy scales of the theory.
For the particle physics phenomenology we focus on the potentials that allow for a 
 stable static background\footnote{Some theoretical arguments in its favor were given
in  \cite{Shaposhnikov:2008xb,Shaposhnikov:2008xi} and we have nothing to add at the
moment.}, which in particular requires $\L_0=0$.We find that the particle spectrum around such a background necessarily contains a massless scalar excitation, the Goldstone boson of the spontaneously-broken scale symmetry. Since this particle can not play the role of the SM Higgs field, we include an additional scalar field (in a realistic theory this can be the complex scalar doublet of the SM.)
After passing to the Einstein frame, we identify the massless field (dilaton)
and the potentially massive field (Higgs boson).\footnote{The Einstein frame is defined as the frame in which the gravitational part of the action is given by the standard Einstein-Hilbert action.} In the Einstein frame the original scale invariance,
existing in the Jordan frame, is replaced by a shift symmetry of the dilaton field. As long as $\L_0=0$, the shift symmetry is unbroken and the dilaton couples only derivatively.
 Hence, it easily avoids experimental bounds on the existence of a
 long-ranged 5th force. The case $\Lambda_0\neq 0$ is
relevant for cosmological considerations (which are deferred to Section \ref{cosmo}).
Besides, for $\Lambda_0\neq 0$ the shift symmetry is broken by the presence of a new interaction term
between the dilaton and the Higgs field with an interaction scale related to the integration constant.

In Sections \ref{ctb} and \ref{ctm} we include gauge
fields and fermions in our considerations and define the conditions yielding
a model with massive gauge vectors (potentials
related to the Higgs model). These results are used in 
 Section \ref{asm} to outline how the new framework can be
applied to the Standard Model. In Section \ref{potentials} we
summarize the requirements to be imposed on the scale-invariant TDiff
Lagrangians which lead to an acceptable low-energy phenomenology.
In addition, we present particular examples of scale-invariant TDiff theories that satisfy these requirements. One of the examples corresponds to the model of \cite{Shaposhnikov:2008xb}.

Section \ref{cosmo} briefly discusses the case $\Lambda_0\neq 0$ and
 cosmological applications.  We will show that for a certain class of potentials, the
 cosmological solutions are very similar to those found in the particular case discussed in \cite{Shaposhnikov:2008xb}.
Namely, the {\em spontaneous} breakdown of  scale invariance due to the flat direction
in the scalar potential dynamically generates Newton's
gravitational constant and particle masses and thereby provides a mechanism for inflation, whereas the breakdown of scale invariance due to the $\Lambda_0$-term leads to dynamical dark energy.
We show here that in spite of the fact that $\Lambda_0 \neq 0$, the
dilaton practically decouples and thus evades all the constraints on extra forces.

We present our conclusions in  Section \ref{concl}.

%%%%%%%%%%%%%%%%%%%%%%%%%%%%%%%%%%%%%%
\section{TDiff invariant theories of gravity}
\label{tdit}
%%%%%%%%%%%%%%%%%%%%%%%%%%%%%%%%%%%%%%

The group of invariance of Einstein's theory of General Relativity is the group of all diffeomorphisms (coordinate changes)
\begin{equation}
x^\m\mapsto \tilde x^\mu(x)\;,
\end{equation}
whose infinitesimal form is
\begin{equation}
\label{eq:Diff}
x^\mu\mapsto x^\mu+\xi^\mu(x)\;.
\end{equation}
We will refer to the group of all diffeomorphisms with the abbreviation Diff.
If gravity is described by a symmetric metric $g_{\mu\nu}$, Diff invariance, together with the requirement that the field equations should contain no higher than second derivatives, uniquely fixes the form of the gravitational action. 
 Diff invariance also dictates how matter fields are coupled to gravity (with the possibility of non-minimal couplings), resulting
 in an extremely successful theory  \cite{Will:2005va}.

Looking for theoretical alternatives to GR, one can consider the question of finding the minimal group of gauge invariance
 giving rise to a satisfactory theory of gravitation (in particular including spin-2 excitations)\footnote{Here we will only consider the case of Lorentz invariant
theories. For
viable theories of gravity without Lorentz invariance  see e.g. \cite{Blas:2010hb,Rubakov:2008nh}.}. The 
answer to this question is the TDiff group that was introduced in (\ref{TD}), cf. \cite{vanderBij:1981ym,Alvarez:2006uu}. This is one of the motivations for exploring TDiff gravity, see e.g. \cite{Buchmuller:1988wx,Alvarez:2006uu,Pirogov:2006zd,Alvarez:2010cg}.

Unlike Diff invariance, TDiff invariance does not uniquely fix the form of the gravitational action. In particular, the action can contain arbitrary functions of the metric determinant, $g$, since it is a scalar under TDiff. The most general TDiff invariant action for gravity
containing no higher than second derivatives is therefore\footnote{We will follow Weinberg's conventions:
$\eta_{\m\n}= \mathrm{diag}(-1,1,1,1)$, 
$R^\a_{\phantom{\a}\b\gamma\delta}=\pd_\delta\Gamma^{\a}_{\beta\gamma}
+\Gamma_{\b\g}^\l\Gamma_{\l\delta}^\a-\;(\g\leftrightarrow\delta)$. Finally $R_{\m\n}=R^\a_{\phantom{\a}\m\a\n}$. In this conventions, 
if $\tilde g_{\mn}=\Omega^2 g_{\m\n}$, $$\tilde R=\Omega^{-2}\left(R+\frac{3}{\sqrt{-g}}\, \pd_\m \left(\sqrt{-g}\,g^{\m\n}\pd_\n \ln\Omega^2\right)
+\frac{3}{2}\, g^{\m\n}\pd_\m \ln\Omega^2\pd_\n \ln \Omega^2\right).$$}
\begin{equation}\label{fulltdlag}
\mathcal{S}_{TD}=\int \di x^4 \sqrt{-g}\left(-\frac{1}{2}M^2f(-g)R-\frac{1}{2}
M^2{\mathcal G}(-g)g^{\mu\nu} \partial_\mu g\partial_\nu g -M^4v(-g)\right)\;,
\end{equation}
where $f(-g)$, ${\mathcal G}(-g)$ and $v(-g)$ are arbitrary functions and $M$ is an a priori arbitrary mass scale. For the previous action to be defined (in particular for the existence of $g^{\m\n}$), it is necessary that $-g>0$, which we will assume henceforth. The couplings between gravity and matter based on TDiff invariance are also much less restricted than in the case of Diff invariance. Just like the gravitational part of the action, they can contain arbitrary functions of $g$. We will refer to the arbitrary functions of $g$ as ``Theory-Defining Functions'' (TDF). Ultimately, all TDF will be restricted by theoretical and phenomenological considerations.

The action $S_{TD}$ describes in general three propagating degrees of freedom, the graviton plus a new scalar. There are two particular choices for the arbitrary functions that enhance the TDiff invariance by an additional local invariance such that the scalar degree of freedom is absent \cite{Alvarez:2006uu}. The first one obviously corresponds to GR ($f=const.,~v=const.,~\mathcal G=0$). The second one corresponds to choosing the functions such that the action is invariant under local (Weyl) rescalings of the metric $g_{\mu\nu}\mapsto e^{2\sigma(x)}g_{\mu\nu}$, where $\sigma(x)$ is an arbitrary function ($f=(-g)^{-1/4},~v=0,~\mathcal G=-\frac{3}{32}(-g)^{-9/4}$). In this second case (sometimes called WTDiff), the action depends on the metric only through the unimodular metric $\hat g_{\mu\nu}=g_{\mu\nu}(-g)^{-1/4}$. Therefore, this case exactly corresponds to UG.

Except for the previous cases, we expect the theory to have arbitrary  $\mathcal G$ and $v$, which implies the existence of a new scalar degree of freedom in the field $g_{\m\n}$. Depending on its mass this 
will have different phenomenological consequences (in particular for searches of 5th forces).  As we will explicitly show in the next section, the theory can be reformulated in the more standard framework where the 
 additional degree of freedom appears as a new type of ``matter'' (or source for the standard GR metric) that can mediate interactions between other fields of the SM. Thus, the distinction between gravity and matter becomes ambiguous in these theories. In particular, this allows us to relate the Higgs field of the SM to the determinant of the metric, and to interpret the  ``new'' interactions within the SM framework.

%%%%%%%%%%%%%%%%%%%%%%%%%%%%%%%%%%%%%%
\subsection{Equivalent Diff invariant theories}
\label{sec_Diff}
%%%%%%%%%%%%%%%%%%%%%%%%%%%%%%%%%%%%%%

It proves very convenient to reformulate TDiff invariant theories as
Diff invariant theories, where the extra degree of freedom appears
explicitly.
In this section we will make use of the St\"uckelberg formalism
to achieve this goal  (see also \cite{Buchmuller:1988wx,Henneaux:1989zc,Alvarez:2007nn,LopezVillarejo:2010iq} for
related works).

Let us consider the generic TDiff invariant Lagrangian (\ref{fulltdlag}).
To start with, note that one can always add an arbitrary constant $\Lambda_0$ to this Lagrangian, without changing the theory. 
Next, one can transform the associated action to an arbitrary coordinate frame by performing a generic Diff transformation. The resulting action
is
\begin{equation}
\label{eq:stuck}
\begin{split}
 \mathcal{S}_e=\int \di^4 x\sqrt{-g}\Bigg(&-\frac{1}{2}M^2f(-g/a)
R-\frac{1}{2}M^2\mathcal G(-g/a)g^{
\m\n}\partial_\m(-g/a)\partial_\n(-g/a)\\
&-M^4v(-g/a)
-\frac{\Lambda_0}{\sqrt{-g/a}}\Bigg)\;,
\end{split}
\end{equation} 
where  $a(x)\equiv J(x)^{-2}$, $J(x)$ being the Jacobian of the Diff {transformation}, and
 $\Lambda_0$ is the aforementioned \emph{arbitrary} constant. The action \eqref{eq:stuck} is classically equivalent
to \eqref{fulltdlag} and the equations of motion for $g_{\m\n}$ are identical. 
Let us now promote $a(x)$ to a dynamical field (commonly called St\"uckelberg, Goldstone or Compensator field) and 
let it transform under the Diff (\ref{eq:Diff}) like the determinant of the metric, i.e. as
\be
\delta_\xi a=\xi^\m\pd_\m a+2a \pd_\m \xi^\m .
\ee
 As a consequence, the action
(\ref{eq:stuck}) is  invariant under Diff,  
\be
\int \di^4 y\left(\frac{\delta \mathcal S_e}{\delta a(y)}\delta_\xi
a(y)+\frac{\delta \mathcal S_e}{\delta g_{\m\n}(y)}\delta_\xi 
g_{\m\n}(y)\right)=0,
\ee
where 
\be
 \delta_\xi
g_{\m\n}=\nabla_{\m}\xi_{\n}+\nabla_{\n}\xi_{\m}.
\ee
If the metric satisfies its equations of motion, the previous
identity is reduced to 
\be
\label{eq:bianhi}
\int \di^4 y \,\sqrt{a}\, \xi^\m\pd_\m\left(\sqrt{a}\frac{\delta \mathcal S_e}{\delta a}\right)=0.
\ee
This identity is valid for arbitrary $\xi^\mu$ and hence
the equations of motion of the metric imply
\begin{equation}
\frac{\delta \mathcal S_e}{\delta a}=\frac{C_0}{\sqrt{a}}\;,
\end{equation}
where $C_0$ is an arbitrary integration constant.
The left-hand side of these equations contains a term proportional to $\L_0$, which has exactly the same form as the term on the right-hand side. Hence, the term of the right-hand side can always be absorbed by a redefinition of the arbitrary constant $\L_0$, resulting in
\begin{equation}
\frac{\delta \mathcal S_e}{\delta a}=0\;.
\end{equation}
This is enough to prove that the equations of motion derived from
\eqref{eq:stuck}, considering $g_{\m\n}$ and $a$ as independent
fields, are equivalent to those derived from \eqref{fulltdlag}, where only $g_{\m\n}$ is varied. By
construction, the new action has an additional local
 invariance. In the gauge $a=1$ (which we assume to be achievable) the solutions of the new
equations are exactly the same as those gotten from \eqref{fulltdlag}.
Solutions derived in a gauge $a\neq 1$ also correspond to the
solutions of \eqref{fulltdlag}, however now written in different
coordinates.

We will refer to the model characterized by the Lagrangian density $\mathcal{L}_e$ in
(\ref{eq:stuck}) as the equivalent Diff invariant theory. Also in
the rest of this paper the subscript $e$ will be used in this sense.
Let us now define the field
\begin{equation}\label{sbiggerzero}
\s\equiv-g/a>0\;,
\end{equation}
which is a scalar under all
diffeomorphisms, and rewrite the Lagrangian as
\begin{equation}
 \mathcal{L}_e=\sqrt{-g}\left(-\frac{1}{2}M^2f(\s)
R-\frac{1}{2}M^2{\mathcal G}(\s)g^{\m\n}\partial_\m\s\partial_\n\s\\
-M^4v_{\Lambda_0}(\s)\label{ediff}\right)\;,
\end{equation} 
where
$$v_{\Lambda_0}(\s)=M^4v(\s)+\frac{\Lambda_0}{\sqrt{\s}}.$$ 
The theory formulated this way reduces to \eqref{fulltdlag} after imposing the gauge
condition  $-g=\s$ (corresponding to $a=1$).  For any other
gauge conditions with $-g\neq \s$ (which may be more convenient for
other reasons), it still corresponds to the original TDiff theory but
written in new coordinates related to the original ones by a
transformation with Jacobian $J\neq 1$.

The appearance of a new parameter $\Lambda_0$ is a general feature of
TDiff theories (see also \cite{Guendelman:2006wr} for other 
{theories of gravity involving} arbitrary integration constants). For $f(-g)=(-g)^{-1/4}*cst.$ (like in pure UG \cite{vanderBij:1981ym,Wilczek:1983as,Zee:1983jg,Henneaux:1989zc}) it plays the role of a cosmological constant.
In all other cases, $\Lambda_0$ leads to a new specific potential term for the scalar field $\sigma$.
At this point, we would like to stress that $\Lambda_0$ is a parameter
characterizing the solution of the equations of motion and is not a
fundamental coupling constant in the action
\eqref{fulltdlag}.  At the classical level, 
$\L_0$ should be understood
as an additional \emph{global} degree of freedom which turns out to be a
constant of the motion: once the initial conditions \emph{and} $\Lambda_0$ are chosen, the evolution
proceeds identically in both, scalar-tensor theories of gravity and TDiff gravity\footnote{In this sense, the
equations of motion of the single
Lagrangian \eqref{fulltdlag} correspond to those of a whole family of
Lagrangians \eqref{ediff} with different values of $\Lambda_0$.}.

 Quantum mechanically, the relation between both 
theories is more subtle (see e.g. \cite{Unruh:1988in,Henneaux:1989zc,Fiol:2008vk,Ng:1990rw}). 
Being  a global degree of freedom, $\Lambda_0$ can be treated in two different ways in the quantum theory. 
First, one can consider the \emph{projected} case, where $\Lambda_0$ is fixed to a certain value. This case
is identical to GR \cite{Fiol:2008vk}. One could also consider $\Lambda_0$ as an integration variable in the path integral formulation of the
theory. However, in the absence of a well-defined path integral formulation of the theory, the results of this approach, though very interesting, should be considered as preliminary 
(see, e.g., \cite{Ng:1990rw,Barrow:2010xt}).

%%%%%%%%%%%%%%%%%%%%%%%%%%%%%%%%%%%%%%
\subsection{Classical backgrounds and local degrees of freedom}
\label{deof}
%%%%%%%%%%%%%%%%%%%%%%%%%%%%%%%%%%%%%%

In this section, we consider the maximally symmetric background solutions of the theory described by
(\ref{ediff}) and determine the conditions, under which they are perturbatively stable.
By a maximally symmetric background solution we mean a solution of the classical equations of motion, which corresponds to constant fields in the particle physics sector of the theory and a maximally symmetric geometry, i.e. Minkowski (flat), de Sitter (dS) or Anti de Sitter (AdS) space-time. The existence of such a ground state may
be essential for a consistent quantization of the theory.
In order to find such solutions it is convenient to first rewrite the theory in the Einstein frame (E-frame), where
the scalar field $\s$ is minimally coupled to the metric. We
define the E-frame metric as
\renewcommand\arraystretch{1.3}
\begin{equation}\label{conf}
\begin{array}{lcl}
\tilde{g}_{\mu\nu}&=&\Omega^{2}g_{\mu\nu},\\
\tilde{g}^{\mu\nu}&=&\Omega^{-2}g^{\mu\nu},\\
\Omega^2&=&f(\sigma)\;,
\end{array}
\end{equation} 
in terms of which the Lagrangian \eqref{ediff} reads
\begin{equation}\label{ediff_ef}
 \mathcal{L}_e=\sqrt{-\tilde{g}}\left(-\frac{1}{2}
M^2\tilde{R}-\frac{1}{2}\mathcal K(\s)M^2\tilde{g}^{\m\n}
\partial_\m\s\partial_\n\s-V_{\Lambda_0}(\s)\right)\; ,
\end{equation} 
where
\begin{align}
 &\mathcal K(\s)=\frac{\mathcal G(\s)}{f(\s)}+\frac{3}{2}\left(\frac{f'(\s)}{f(\s)}
\right)^2\; , \quad
 V_{\Lambda_0}(\s)=\frac{v_{\Lambda_0}(\s)}{f(\s)^2}
\label{AD}\;.
\end{align} 
For the previous transformation to make sense as a field redefinition between
two field theories defined perturbatively around a certain 
background $\s_0$, one should assume
\be
\label{cond2}
f(\s_0)\neq 0.
\ee
This is at the same time the condition leading to an induced gravitational scale
and weakly interacting spin-two excitations around this background (cf. (\eqref{ediff}). We will assume it
henceforth.

For a constant scalar field $\s=\s_0$, the equations of motion imply
\begin{equation}\begin{array}{lcl}
\Lambda_0&=&2M^4\s_0^{3/2}\left(\dfrac{v'(\s_0)f(\s_0)-2f'(\s_0)v(\s_0)}{f(\s_0)+4\s_0f'(\s_0)}\right),\\[0.5cm]
\tilde R&=&-4M^2\left(\dfrac{v(\s_0)+2\s_0v'(\s_0)}{f(\s_0)[f(\s_0)+4\s_0f'(\s_0)]}\right).
\end{array}
\end{equation}
Unless the right-hand side vanishes, the first equation can be understood as an equation for $\s_0$ in terms of $\L_0$.
The second equation shows  that the solution may be flat, dS or AdS, depending on the TDF (and on $\L_0$ through $\s_0$). 
In the degenerate case $f(\s_0)+4\s_0f'(\s_0)=0$ (which, in particular, corresponds to UG), a maximally symmetric background solution only exists, if there is a value $\s_0$, for which $v(\s_0)+2\s_0v'(\s_0)=0$.
The classical ground state is then given by $\s=\s_0$ and $\tilde R=-4V_{\L_0}M^{-2}$. Again, depending on the TDF, respectively the value of $\L_0$, the corresponding maximally symmetric space-time is flat, dS or AdS.

Thus,  in a TDiff theory containing the metric as the only field, maximally symmetric background solutions always exist, independently of the TDF (except may be in the degenerate case). Flat space-time is a solution, provided
\begin{equation}\label{flatcond}
v(\s_0)+2\s_0v'(\s_0)=0\;.
\end{equation}

For the study of the propagating degrees of freedom we will focus on the case in which  flat space-time is a solution, 
\be
\label{cs}
\tilde g_{\m\n}=\eta_{\m\n}, \quad \s=\s_0, \quad \Lambda_0=-M^4v(\s_0)\sqrt{\s_0},
\ee
and introduce the perturbations
\begin{align}
\label{perts}
&\tilde{g}_{\mu\nu}=\eta_{\mu\nu}+\frac{\tilde{h}_{\mu\nu}}{M}\;,\quad 
\sigma=\sigma_0+\frac{\varsigma}{M}\;.
\end{align}
The  part of the  Lagrangian \eqref{ediff_ef} quadratic in
perturbations reads\footnote{Here and in the rest of the paper, we use
the notation $F^{(n)}=\frac{d^n F(\s)}{d\s^n}\Big|_{\s=\s0}$ for the
derivatives of functions evaluated at the background field value .}
\begin{equation}
\label{tdquad}
 \mathcal{L}_e^{Q}=\frac{1}{2}\tilde{\mathcal{L}}_{EH}^{Q}-\frac{1}{
2}
\mathcal K^{(0)}\eta^{\mu\nu}
\partial_\mu \varsigma\partial_\nu
\varsigma-\frac{1}{2}V_{\Lambda_0}^{(2)}M^{-2}\varsigma^2\;,
\end{equation}
where the first term in \eqref{tdquad} is the standard quadratic
Einstein-Hilbert Lagrangian
\be
\tilde{\mathcal{L}}_{EH}^{Q}=-\frac{1}{4}\partial_\r
\tilde h_{\m\n}\partial^\r \tilde h^{\m\n}+\frac{1}{2}\partial^\n
\tilde h_{\m\n}\partial^\r \tilde h^\m_{\phantom{\s}\r}-\frac{1}{2}\partial_\m
\tilde h\partial_\n \tilde h^{\m\n}+\frac{1}{4}\partial_\m
\tilde h\partial^\m \tilde h\;,
\ee
with indices raised and lowered with the Minkowski metric
$\eta_{\m\n}$ and $\tilde h\equiv \tilde h_{\phantom{\m}\m}^\m$.  This term
describes two massless tensor degrees of freedom. From \eqref{tdquad}
one can see that whenever $\mathcal K^{(0)}=\mathcal K(\sigma_0)\neq 0$, the theory also
contains a scalar degree of freedom\footnote{Both for GR  ($f(\s)=1$ and $\mathcal G(\s)=v(\s)=0$) and for UG ($f=\s^{-1/4},~{\mathcal G}=-\frac{3}{32}\s^{-9/4},~v=0$) one finds $\mathcal K^{(0)}=0$
and hence these theories only contain the two massless tensor degrees of
freedom.}. In that case, the scalar part can be brought to canonical
form by defining the canonically normalized  field
\begin{equation}
\varsigma_c=\sqrt{\left|\mathcal K^{(0)}\right|}\varsigma\;.
\end{equation}
We get
\begin{equation}
 \mathcal{L}_e^{Q}=\frac{1}{2}\tilde{\mathcal{L}}_{EH}^{Q}
-\epsilon_\varsigma\frac{1}{2}\partial_\mu
\varsigma_c\partial^\mu \varsigma_c
-\frac{m_\varsigma^2}{2}\varsigma_c^2\;,
\end{equation}
where
\begin{align}
\label{mass}
 &\epsilon_\varsigma\equiv
\textnormal{sign}\left(\mathcal K^{(0)}\right),\quad 
 m_\varsigma^2\equiv\epsilon_\varsigma\frac{V_{\Lambda_0}^{(2)}}{\mathcal K^{(0)}}
M^{-2}\;.
\end{align}
The perturbations around the background (\ref{cs}) will be well-behaved
provided that:
\begin{itemize}
 \item The scalar field $\varsigma_c$ has a positive definite kinetic
term (absence of ghosts): $\mathcal K^{(0)}>0$.
\item  The field $\varsigma_c$  has positive or zero mass (absence of
tachyons): $V_{\Lambda_0}^{(2)}\geq 0$.
\end{itemize}
Finally, on top of the terms quadratic in the perturbations there is obviously
a series of interaction terms. We will get interested in those terms
in the upcoming sections, where we consider the phenomenology of different types of fields
coupled to TDiff gravity.

%%%%%%%%%%%%%%%%%%%%%%%%%%%%%%%%%%%%%%
\section{Scale-invariant TDiff theories}
\label{sitd}
%%%%%%%%%%%%%%%%%%%%%%%%%%%%%%%%%%%%%%
In this section we will focus on scale-invariant TDiff theories
including scalar fields only. Other SM fields will be
introduced in the subsequent sections.

Assuming that the metric  has a
non-zero scaling dimension $d_g\neq 0$, the Lagrangian (\ref{fulltdlag}) is invariant under the scale transformation
\be
\label{scaleg}
g_{\m\n}(x)\mapsto \l^{d_g}g_{\m\n}(\l x),
\ee
provided that the TDF satisfy, 
\be
\label{TDF_s}
f(-g)=f_0\,(-g)^{\frac{2-d_g}{4d_g}}, \quad \mathcal{G}(-g)=\mathcal{G}_0\,(-g)^{-2+\frac{2-d_g}{4d_g}}, \quad v(-g)=v_0\,(-g)^{\frac{2-d_g}{2d_g}}\;,
\ee
where $f_0$, $\mathcal G_0$ and $v_0$ are arbitrary constants.

The scaling dimension of the metric can be changed by performing the TDiff compatible field redefinition
($\a$ is a real constant)
\begin{equation}\label{mrd}
g_{\mu\nu}\mapsto (-g)^{\alpha}g_{\mu\nu}.
\end{equation}
Thus, different scaling dimensions correspond to equivalent theories. 
In particular, we could set $d_g=2$. The scale transformation with $d_g=2$ corresponds to a diffeomorphism and hence any Diff invariant theory is 
  scale-invariant in the aforementioned sense (notice that the opposite is not true: not any scale-invariant TDiff theory is Diff invariant).
In particular, GR corresponds to the Lagrangians with TDF \eqref{TDF_s} satisfying $\frac{\mathcal{G}_0}{f_0}=-\frac{3}{2}\left(\frac{2-d_g}{4d_g}\right)^2$ and $v_0=0$. Let us mention that
UG is not invariant under \eqref{scaleg}.

Following the discussion in Section \ref{deof}, we may look for maximally symmetric solutions for the TDF (\ref{TDF_s}). 
Recalling that $\sigma>0$,  one finds that $f(\s_0)+4\s_0f'(\s_0)\neq 0$ for all values of
$\sigma_0$. Hence, maximally symmetric background solutions always exist.
These solutions spontaneously break the scale symmetry.
The condition \eqref{flatcond} for the existence of the flat space-time solution is fulfilled only if $v_0=0$. 
Except for the case corresponding to GR, the spectrum around the flat background solution contains a propagating massless scalar degree of freedom. It represents the Goldstone boson of the spontaneously broken scale invariance.
 
The above theory is interesting because of its uniqueness. However, it is not enough for our purposes, since we want to build a theory containing a massive scalar field that plays the role of the SM Higgs field. In view of this, we will now consider the possibility of adding
an extra real scalar field, $\phi$. A scale-invariant TDiff theory including $g_{\mu\nu}$ and $\phi$ will be invariant under the transformations
\begin{equation}
g_{\m\n}(x)\mapsto\lambda^{d_g}g_{\m\n}(\lambda x),\quad \phi(x)\mapsto \lambda^{d_\phi}\phi(\lambda x)\;.
\end{equation}
By a field redefinition of the type
\begin{equation}
g_{\m\n}\mapsto (-g)^\a\phi^\b g_{\m\n},\quad \phi\mapsto (-g)^\g\phi^\d\;,
\end{equation}
compatible with the TDiff invariance, the scaling dimensions of the fields can always be changed to different values. In other words, the way one attributes scaling dimensions to the different fields merely corresponds to the choice of field variables. Without loss
of generality, we will choose the scaling dimensions to correspond to the usual canonical mass dimensions, i.e. $d_g=0$ and $d_\phi=1$, for which the scale transformation is
\begin{align}
\label{st}
 g_{\m\n}(x)\mapsto
g_{\m\n}(\lambda x),\quad \phi(x)\mapsto \lambda\phi(\lambda x) \;.
\end{align}
An alternative choice of the scaling dimensions would be $d_g=2$ and $d_\phi=0$. In this case, the scaling dimensions of the fields correspond  to their tensorial rank. This choice reveals an interesting property of scale-invariant TDiff theories: the group of invariance including TDiff and the scale transformations where all fields have scaling dimension equal to their tensorial rank can be identified as a subgroup of Diff (see also
the comments after Eq.~\eqref{mrd}). In other words, the symmetry group consisting of TDiff plus global scale transformations constitutes a subgroup of the full Diff group.

Supplementing the model \eqref{fulltdlag} by a real scalar field and imposing invariance under the transformation \eqref{st}, one finds
the action\footnote{Terms with arbitrarily many derivatives can be included in a scale-invariant way, if one allows for $\phi$ to appear in the denominator.
We will assume that, if present, these terms are suppressed by a large energy scale.}
\be
\begin{split}
\label{fulltdscale}
\mathcal S=\int \di x^4{\sqrt{-g}}\Big(-\frac{1}{2}\phi^2f(-g)R-&\frac{1}{2}
\phi^2\mathcal{G}_{gg}(-g)(\partial g)^2
-\frac{1}{2}\mathcal{G}_{\phi\phi}(-g)(\partial\phi)^2\\
&+\mathcal{G}_{g\phi}(-g)\phi
\,\partial g\cdot \partial\phi-\phi^4v(-g)\Big)\;.
\end{split}
\ee
Here, and in many of the upcoming expressions, in order to shorten
notations, we do no longer write Lorentz indices explicitly. The
implicit contractions of Lorentz indices are done with the metric
$g_{\m\n}$ if the Lagrangian is in the J-frame and with $\tilde
g_{\m\n}$ if it is in the E-frame. Like in the theory containing only the metric field (Section \ref{tdit}), the Lagrangian \eqref{fulltdscale} contains arbitrary functions (TDF) of the metric determinant $g$. The dependence on the scalar field $\phi$ is dictated by scale invariance. Note, however, that the situation is different if one chooses variables such that $d_g=2$ and $d_\phi=0$. In that case, scale invariance dictates the dependence of the Lagrangian on $g$, while the arbitrary functions solely depend on $\phi$.

Using the St\"uckelberg formalism illustrated in Section
\ref{sec_Diff} we can write down the Lagrangian of the equivalent Diff invariant theory
of \eqref{fulltdscale} as
\be
\begin{split}
\frac{\mathcal{L}_e}{\sqrt{-g}}=
-\frac{1}{2}\phi^2f(\s)R-&\frac{1}{2}\phi^2\mathcal{G}_{gg}
(\s)(\partial \s)^2
-\frac{1}{2}\mathcal{G}_{\phi\phi}(\s)(\partial\phi)^2\\
&
-\mathcal{G}_{g\phi}(\s)\phi~\partial\s\cdot\partial\phi-\phi^4v(\s)
-\frac{\Lambda_0}{\sqrt{\s}}\;.
\label{fulltdscale_diff}
\end{split}
\ee
For $\L_0=0$, the corresponding action is invariant under \eqref{st} supplemented by the transformation of $\s$, i.e.
\begin{align}
\label{stsupp}
 g_{\m\n}(x)\mapsto
g_{\m\n}(\lambda x),\quad \phi(x)\mapsto \lambda\phi(\lambda x),\quad \s(x)\mapsto\s(\lambda x) \;.
\end{align}
In fact, in this case,  \eqref{fulltdscale_diff} is also invariant under  the internal transformation
\begin{align}
\label{stint}
 g_{\m\n}(x)\mapsto
\l^2 g_{\m\n}(x),\quad \phi(x)\mapsto \lambda^{-1}\phi(x) \;.
\end{align}
A non-zero $\Lambda_0$ breaks these symmetries. Thus, scale-invariant TDiff theories naturally produce a unique symmetry-breaking potential term. This should be contrasted with the situation in generic Diff invariant theories, where such a term could only be introduced ad hoc. In other words, starting from a Diff invariant theory, there would be no reason to include in (\ref{fulltdscale_diff}) the term proportional to $\L_0$  without also including all other possible terms breaking the scale symmetry (\ref{stsupp}).

The Lagrangian (\ref{fulltdscale_diff}) can be transformed to
the Einstein
frame (provided $\phi^2f(\s) \neq 0$), with the help of a conformal
transformation
\begin{align}
\label{eq:Eins_scale}
 &\tilde{g}_{\mu\nu}=\Omega^{2}g_{\mu\nu}, 
 \quad \tilde{g}^{\mu\nu}=\Omega^{-2}g^{\mu\nu}\;,
 \quad \Omega^2=\frac{\phi^2f(\s)}{M^2}\;.
\end{align} 
It takes the form
\be
\begin{split}
\frac{\mathcal{L}_e}{\sqrt{-\tilde{g}}}=
-\frac{1}{2}M^2\tilde{R}&-\frac{1}{2}M^2\mathcal{K}_{\s\s}(\s)(\partial\s)^2
-\frac{1}{2}M^2\mathcal{K}_
{\phi\phi}(\s)(\partial\ln(\phi/M))^2\\
&-M^2\mathcal{K}_{\s\phi}(\s)~\partial
\s\cdot\partial\ln(\phi/M)-M^4V(\s)-\frac{M^4\Lambda_0}{\phi^4f(\s)^2\sqrt{\s}}
\;,
\label{fulltdscale_diff_einI}
\end{split} 
\ee
where
\renewcommand\arraystretch{2}
\begin{equation}\begin{array}{lclclcl}
\mathcal{K}_{\s\s}(\s)&=&\dfrac{\mathcal{G}_{gg}(\s)}{f(\s)}+\dfrac{3}{2}\left(\dfrac{
f'(\s)}{f(\s)}\right)^2&, 
&\quad \mathcal{K}_{\phi\phi}(\s)
&=&\dfrac{\mathcal{G}_{\phi\phi}(\s)}{f(\s)}+6\;, \\
 \mathcal{K}_{\s\phi}(\s)
 &=&\dfrac{\mathcal{G}_{g\phi}(\s)}{f(\s)}+3\dfrac{f'(\s)}
{f(\s)}&,&\quad 
V(\s)&=&\dfrac{v(\s)}{f(\s)^2}\;.\label{BC}
\end{array}\end{equation}
Except for the term proportional to $\L_0$, the E-frame action is invariant under scale transformations with $d_{\tilde g}=2$ and $d_\phi=1$.\footnote{Note that in the equivalent Diff invariant formulation, the action is exactly invariant under scale transformations with $d_g=d_{\tilde g}=2$ and $d_\phi=0$ as these transformations are a part of Diff.} That is why, in the scale-invariant part, $\phi$ can only enter in the combination
\be
\label{scal_inv_op}
\pd_\m \ln (\phi/M).
\ee
This can also be understood from the fact that in the E-frame the transformation \eqref{stint} becomes
\begin{align}
\label{stintE}
 \tilde g_{\m\n}(x)\mapsto
\tilde g_{\m\n}(x),\quad \phi(x)\mapsto \lambda^{-1}\phi(x) \;.
\end{align}
The kinetic term for the scalar fields can be diagonalized by
redefining the fields as\footnote{Here we assume that both
$\mathcal{K}_{\phi\phi}(\s)$ and
$\frac{\mathcal{K}_{\s\s}(\s)\mathcal{K}_{\phi\phi}(\s)-\mathcal{K}_{
\s\phi}
(\s)^2}{\mathcal{K}_{\phi\phi}(\s)}$ are non-zero.}
\begin{align}
 \tilde{\s}=\int_{\s_0}^\s
\di\s'\sqrt{\left|\frac{\mathcal{K}_{\s\s}(\s')\mathcal{K}_{\phi\phi}(\s')
-\mathcal{K}_{\s\phi}
(\s')^2}{\mathcal{K}_{\phi\phi}(\s')}\right|},
 \quad \tilde{\phi}=M\left(\ln\frac{\phi}{M}+\int_{\s_0}^\s
\di\s'\frac{\mathcal{K}_{\s\phi}(\s')}{\mathcal{K}_{\phi\phi}(\s')}\right)
\label{tr0}.
\end{align}
Note that we chose the integration constant such that $\tilde
\sigma(\sigma_0)=0$ and kept $\sigma_0$ arbitrary for the moment.
After
this field redefinition (which is always solvable in perturbation
theory) the Lagrangian simplifies to
\be
\frac{\mathcal{L}_e}{\sqrt{-\tilde{g}}}=
-\frac{1}{2}M^2\tilde{R}-\frac{1}{2}\e_\s M^2
(\partial\tilde\s)^2-\frac{1}{2}\tilde{\mathcal{K}}_{\phi\phi}(\tilde{\s})
(\partial\tilde\phi)^2
-M^4\tilde{V}(\tilde{\s})-\Lambda_0\, \tilde{\mathcal{K}}_{\Lambda_0}(\tilde
\s)\exp\left(-\frac{4\tilde\phi}{M}\right)
\;,
\label{fulltdscale_diff_ein}
\ee
where $\e_{\s}=\textnormal{sign}\left(\frac{\mathcal{K}_{\s\s}(\s)
\mathcal{K}_{\phi\phi} (\s)-\mathcal{K}_{\s\phi}
(\s)^2}{\mathcal{K}_{\phi\phi}(\s)}\right)$, 
and the different functions are obtained by expressing
$\sigma$ as a function of $\tilde \s$,
 \be
 \label{eq:tildepot}
 \tilde V(\tilde \s)=V(\s),\quad \tilde{\mathcal{K}}_{\phi\phi}(\tilde{\s})
 =\mathcal{K}_{\phi\phi}(\s),\quad
 \tilde{\mathcal{K}}_{\Lambda_0}(\tilde \s)=\frac{\exp\left(4\int_{\s_0}^\s
\di\s'\frac{\mathcal K_{\s\phi}(\s')}{\mathcal K_{\phi\phi}(\s')}\right)}
{f(\s)^2\sqrt{
\s}}\;.
\ee
After the field redefinition (\ref{tr0}), the scale transformation for $\phi$ translates into the invariance under
global shifts of the dilaton
field,  $\tilde\phi\mapsto \tilde\phi+\lambda$. This can be understood as the E-frame manifestation
of scale invariance in the J-frame. If $\Lambda_0=0$, the dilaton
is exactly massless, and interacts with the field $\tilde\s$ (matter field) only through
derivatives. In other words, it does not lead to measurable long-range
interactions (for experimental bounds on light dilatons see e.g. \cite{Damour:2010rp,Ellis:1987qs}).

This Lagrangian, when considered at the quantum field theory level,
can be regularized by the standard procedures, such as dimensional or
Pauli-Villars regularization. The subtraction procedure  is then
consistent with the shift symmetry and Diff invariance, i.e. this theory is anomaly
free even if one uses the standard regularization schemes. Transforming the E-frame theory (action (\ref{fulltdscale_diff_ein}) plus counter-terms)
 back to the J-frame will result in a quantum theory with exact scale invariance.
In dimensional regularization the exact invariance will be due to a
field-dependent subtraction point, as described in \cite{Englert:1976ep,Shaposhnikov:2008xi}, while if Pauli-Villars or lattice regularizations are used, it will be due to a field-depend mass, respectively lattice spacing \cite{Shaposhnikov:2008ar}.

%%%%%%%%%%%%%%%%%%%%%%%%%%%%
\subsection{Classical backgrounds and local degrees of freedom}
\label{ldof}
%%%%%%%%%%%%%%%%%%%%%%%%%%%%
In this subsection we perform the analysis of maximally symmetric solutions 
and degrees of freedom for scale-invariant TDiff theories. As in Section \ref{deof}, 
we will perform the analysis in the E-frame and assume that
scale invariance is spontaneously broken; in particular
\be
\label{ssb_sc}
\phi_0^2 f(\s_0)\neq 0.
\ee
Once the previous condition is satisfied, Newton's constant (and other scales of the theory) are induced
by the non-zero value of $\phi_0$.

For maximally symmetric solutions, the scalar fields must be constant. Contrary 
to the previous case (cf. Section \ref{deof}), this automatically sets the constant $\L_0=0$ (other possibilities
relevant for cosmological applications will be considered in Section \ref{cosmo}). After setting $\s=\s_0$, the equation of motion for $\s$ yields the condition
\be
V'(\s_0)=0,
\ee
or, in terms of the original TDF,
\be
\label{minim_cond}
f(\s_0)v'(\s_0)-2f'(\s_0)v(\s_0)=0.
\ee
If this condition holds, the remaining equations for the  background fields are
\be
\label{vacu_scale}
\phi=\phi_0, \quad \tilde R=-4M^2\frac{v(\s_0)}{f(\s_0)^2}, \quad \L_0=0,
\ee
where $\phi_0$ is not fixed by the equations of motion (as a consequence of scale invariance). 
For $v(\s_0)\neq 0$, the background will correspond to a dS or AdS space, while
the Minkowski background is obtained for $v(\s_0)=0$. This, together with the 
constraint (\ref{minim_cond}), implies that in the scale-invariant theory with spontaneous symmetry breaking, the existence 
of a Minkowski background requires (compare with (\ref{flatcond}))
\begin{equation}
\label{maincond}
 v(g_0)=v'(g_0)=0\;.
\end{equation} 
Once these conditions are satisfied, 
the Lagrangian (\ref{fulltdscale_diff_ein})
has a background solution with
\begin{align}
 &\tilde g_{\mu\nu}=\eta_{\mu\nu}\;,\quad
 \tilde\sigma=0,\quad
\tilde \phi=\tilde \phi_0,
 \quad \Lambda_0=0\;,\label{cs2}
\end{align}
where $\tilde\phi_0$ is an arbitrary real constant.
We define the perturbations to the background as
\begin{equation}\label{perts0}\begin{array}{lcl}
\tilde{g}_{\mu\nu}&=&\eta_{\mu\nu}+\dfrac{\tilde{h}_{\mu\nu}}{M}\;,\\
\tilde{\s}&=&\dfrac{\varsigma}{M}\;,\\
\tilde{\phi}&=&\tilde{\phi}_0+\dfrac{\varphi}{\sqrt{\left|\tilde{\mathcal{K}}_{
\phi\phi }^{(0)}\right|}} \; .
\end{array}\end{equation}
In the rest of Section \ref{sitd}, Lorentz indices are raised, lowered and
contracted with the Minkowski metric $\eta_{\m\n}$. Let us split the
Lagrangian into a term quadratic in the perturbations and an
interaction term
\begin{equation}
\mathcal{L}_e=\mathcal{L}_e^Q+
\mathcal{L}_e^{(\textnormal{int})}\;.
\end{equation}
For the quadratic term we get
\begin{equation}
\mathcal{L}_e^Q=\tilde{\mathcal{L}}_{EH}^Q
-\e_\varsigma\frac{1}{2}(\partial\varsigma)^2
-\e_\varphi\frac{1}{2}(\partial\varphi)^2
-\frac{1}{2}m_\varsigma^2\varsigma^2\;,
\end{equation} 
where we have defined
\begin{equation}
\label{mass_sc}
\begin{split}
&\epsilon_\varsigma\equiv
\textnormal{sign}\left(\frac{\mathcal{K}_{\s\s}^{(0)}
\mathcal{K}_{\phi\phi}^{(0)}
-\left(\mathcal{K}_{\s\phi}^{(0)}\right)^2}
{\mathcal{K}_{\phi\phi}^{(0)}}\right),
\quad \epsilon_\varphi\equiv
\textnormal{sign}\left(\mathcal{K}_{\phi\phi}^{(0)}\right)\;,\\
&m_\varsigma^2\equiv\epsilon_\varsigma\tilde{V}^{(2)}M^{2}
=\epsilon_\varsigma\left(\frac{\mathcal{K}_{\s\s}^{(0)}
\mathcal{K}_{\phi\phi}^{(0)}
-\left(\mathcal{K}_{\s\phi}^{(0)}\right)^2}
{\mathcal{K}_{\phi\phi}^{(0)}}\right)^{-1}
V^{(2)}M^{2}\;.
\end{split}
\end{equation}
In this case, on top of the two tensorial massless degrees of freedom,
the theory contains \emph{two} scalar degrees of freedom among which
at least one is massless. We have the following criteria for the
perturbations to be well-behaved:\footnote{These conditions can also
be formulated  in a variable independent way. The first two conditions
correspond to a positive definite field space metric to lowest order
in the expansion around the constant background. Requiring that the
matrix of second derivatives of the potential evaluated at the
constant background solution should have no negative eigenvalue is the
analog of the third condition.}
\begin{itemize}
 \item For positive definite kinetic terms (absence of ghosts):\\
\be\label{kc0} 
\mathcal{K}_{\s\s}^{(0)}\mathcal{K}_{\phi\phi}^{(0)}
-\left(\mathcal{K}_{\s\phi}^{(0)}\right)^2>0\quad\textnormal{and}\quad
\mathcal{K}_{\phi\phi}^{(0)}>0\;.
\ee
\item For positive or zero mass of $\varsigma_c$ (absence of
tachyons):
\be\label{mc0} 
V^{(2)}\geq 0\;.
\ee
\end{itemize}
Besides,  there will be phenomenological constraints coming from
the coupling of the previous fields to other fields of the SM. The
only remark we want to make on this respect  is that the massless field $\varphi$ will
be only derivatively coupled to $\s$ (and, moreover, by higher dimensional operators),  which
implies that its effects at small energies are naturally suppressed
(see Section \ref{intera}).

%%%%%%%%%%%%%%%%%%%%%%%%%%%%
\subsection{Interactions and separation of scales}
\label{intera}
%%%%%%%%%%%%%%%%%%%%%%%%%%%%

We now want to consider the interactions coming from the  Lagrangian
(\ref{fulltdscale_diff_ein}) for $\L_0=0$. In general those are represented by an infinite
series of terms arising from the expansion of the functions
$\tilde{\mathcal{K}}_{\phi\phi}(\tilde{\s})$ and
$\tilde{V}(\tilde{\s})$ and of the metric tensor around the constant
background. The interaction terms obtained from the expansion of the
Ricci scalar in  (\ref{fulltdscale_diff_ein}) are suppressed by
the Planck mass. We neglect them, as we will be interested in 
sub-Planckian processes (we consider the cut-off of the theory to be the mass scale $M$).
 Let us consider the terms of dimension up to
four:
\begin{equation}
 \mathcal{L}_e^{\textnormal{int}\leq 4}=-\frac{1}{3!}\k_\varsigma
\varsigma^3-\frac{\l_\varsigma}{4!}\varsigma^4-\frac{1}{4}
\frac{m_\varsigma^2}{M
} \varsigma^2 \tilde{h}-\frac{1}{16}\frac{m_\varsigma^2}{M^2}
\varsigma^2\left(\tilde{h}^2
-2\tilde{h}_{\m\n}\tilde{h}^{\m\n}\right)-\frac{1}{12}\frac{\k_\varsigma}{M}
\varsigma^3\tilde{h}\;,
\label{dim4}
\end{equation} 
where
\begin{align}
 &\k_\varsigma\equiv\tilde{V}^{(3)}M,\quad\quad
 \l_\varsigma\equiv\tilde{V}^{(4)}\;.
\end{align}
These are the relevant operators for a scalar field minimally coupled
to Einstein gravity. For a generic theory (where the TDF and their
derivatives are of the order of one) the mass of the field 
$\varsigma$ is of the order of the Planck scale (cf. (\ref{mass_sc})). If we want to identify the field $\varsigma$ with a low-energy degree of
freedom (such as the Higgs boson of the SM), the TDF must obey several
constraints. 
In particular, the mass of the particle must be much smaller than the mass scale $M$ (which sets 
the cut-off scale of the theory) :
\begin{align}
 &\left|\frac{m_\varsigma}{M}\right|=\sqrt{\left|\tilde{V}^{(2)}
\right|}\ll 1\;.
\label{deccond0}
\end{align}
This condition is similar to the fine-tuning
conditions of the SM, requiring that the Fermi scale is much smaller
than the Planck scale.

Besides, for the theory to be weakly coupled at  energies of order $m_\varsigma$, we also need to have
$\frac{\kappa_\varsigma}{m_\varsigma},\l_\varsigma\lesssim 1$, which means\begin{equation}\label{wc0}
\tfrac{\left|\tilde{V}^{(3)}\right|}{\sqrt{\left|\tilde{V}^{(2)}
\right|}},\,\tilde{V}^{(4)}\lesssim 1\;.
\end{equation}

For the Lagrangian (\ref{dim4}) to represent a  consistent effective field theory at energies smaller than $M$,  the corrections to it
originating from the power expansions of the TDF must be suppressed (see, however, Section \ref{varchoice}). The
higher dimensional operators can be written schematically as
\begin{align}
\mathcal{L}_e^{\textnormal{int}>4}=\phantom{+}&\sum_{ n_h>0 }
^\infty\frac{1}{M^{n_h}}\left(\mathcal{L}_e^Q+\mathcal{L}_e^{
\textnormal{int}\leq 4}\right)\tilde{h}^{n_h}\nonumber\\
+&\sum_ {\substack {n_h\geq 0\\
n_\varsigma>0}}
^\infty\left(\frac{1}{M_{\phi\phi}(n_h,n_\varsigma)}\right)^{n_h+n_\varsigma}
(\partial\varphi)^2 \tilde{h}^{n_h}\varsigma^{n_\varsigma}
+\sum_{\substack{n_h\geq 0\\n_\varsigma>4}}
^\infty\left(\frac{1}{M_V(n_h,n_\varsigma)}\right)^{n_h+n_\varsigma-4}
\tilde{h}^ {n_h} \varsigma^{ n_\varsigma}\;,\label{intlag0}
\end{align} 
where we neglect numerical factors of order one, neglect tensor 
indices and define
\begin{equation}\begin{array}{lcl}
M_{\phi\phi}(n_h,n_\varsigma)&\sim &M\left|\dfrac
{\tilde{\mathcal{K}}_{\phi\phi}^{(n_\varsigma)}}
{\tilde{\mathcal{K}}_{\phi\phi}^{(0)}}\right|^{\frac{-1}{
n_h+n_\varsigma}}~,\\
 M_V(n_h,n_\varsigma)&\sim &M\left|\tilde{V}^{
(n_\varsigma)}\right|^{\frac{-1}{n_h+n_\varsigma-4}}~.
\end{array}\end{equation}
The first line of \eqref{intlag0} represents the standard higher
dimensional operators for Einstein gravity and a minimally coupled
scalar field. If the conditions \eqref{deccond0} and (\ref{wc0}) hold, they are all
suppressed at energies below the scale $M$. The remaining terms are 
 new higher dimensional operators that appear if the
kinetic term is non-canonical and/or if the potential contains higher
dimensional operators. The suppression scales of these
operators are given by $M_{\phi\phi}(n_h,n_\varsigma)$ and
$M_V(n_h,n_\varsigma)$. They are at least of the order of the
Planck scale $M$ provided that
\begin{equation}\label{supcond0}
 \left|\frac
{\tilde{\mathcal{K}}_{\phi\phi}^{(n_\varsigma)}}
{\tilde{\mathcal{K}}_{\phi\phi}^{(0)}}
\right|^{\frac{1}{
n_h+n_\varsigma}}\leq 1
 \quad\;\textnormal{and}\quad\;
 \left|\tilde{V}^{
(n_\varsigma)}\right|^{\frac{1}{n_h+n_\varsigma-4}}\leq 1\;.
\end{equation}

Let us now summarize the findings of this section. We have considered
a scale-invariant theory of a scalar field coupled to TDiff gravity, 
which is described by the action \eqref{fulltdscale} (or equivalently (\ref{fulltdscale_diff_ein})). If
there exists a value of $\s_0$ for which ${v(\s_0)=v'(\s_0)=0}$ (i.e. ${\tilde V^{(0)}=\tilde
V^{(1)}=0}$), there exists a
family of maximally symmetric solutions of the equations of motion, corresponding to flat space-time and constant scalar fields. Those
solutions for which ${\phi_0\neq 0}$ spontaneously break the
dilatation symmetry of the theory. Besides, scale invariance can be independently broken by  an integration constant $\Lambda_0$, which
introduces a run-away potential for  the dilaton field. The quadratic
analysis of perturbations around the background solutions with
$\Lambda_0=0$ shows that if the conditions ${{\mathcal
K}_{\s\s}^{(0)}{\mathcal K}_{\phi\phi}^{(0)} -\left({\mathcal
K}_{\s\phi}^{(0)}\right)^2>0}$ and ${\tilde{{\mathcal
K}}_{\phi\phi}^{(0)}>0}$ are satisfied, the theory describes two
massless tensor degrees of freedom, a massless scalar and a scalar of
mass ${m_\varsigma^2=\tilde V^{(2)}}$. The scale $M$ for gravity and
the scales $m_\varsigma$ and $\k_\varsigma$ associated to the scalar
field are induced by the non-zero value of $\phi_0$.\footnote{This
fact is easier to see in the J-frame. Expanding around the
constant background one finds that the coupling constant of the tensor
modes as well as the mass of the scalar mode are proportional to
$\phi_0$. In the E-frame this fact is implicit, since the
transformation to the E-frame is only allowed if $\phi_0\neq
0$.} If the theory-defining functions are such that the conditions
\eqref{deccond0}, \eqref{wc0} and \eqref{supcond0} are fulfilled, the
scalar and the tensor sectors decouple and all the non-renormalizable
interactions are suppressed below the scale $M$. In this case, at
energies well below $M$, the scalar field phenomenology resulting
from the theory
\eqref{fulltdscale} is indistinguishable from the phenomenology of the
corresponding renormalizable scalar-field theory.

%%%%%%%%%%%%%%%%%%%%%%%%%%%%%%%%%%%%%%
\subsection{Dependence on the choice of variables and  exact renormalizability}
\label{varchoice}
%%%%%%%%%%%%%%%%%%%%%%%%%%%%%%%%%%%%%%

Under very general assumptions, the Lagrangian \eqref{fulltdscale} 
can be brought to the form \eqref{fulltdscale_diff_ein} by a 
non-singular change of variables. Furthermore, one may still perform
 field  redefinitions of the form ${(\tilde
\s, \tilde \phi)\mapsto (\tilde \s', \tilde \phi')}$ that modify the explicit expressions of the functions $\tilde{\mathcal
K}_{\phi\phi}(\tilde\s)$, $\tilde V(\tilde\s)$, etc. For example, for some functions $\tilde {\mathcal
K}_{\phi\phi}(\tilde\s)$ one can make a change of variables  which brings the
kinetic term to the canonical form (see below). Also the functions
$\tilde V(\tilde \s)$ and $\tilde {\mathcal K}_{\L_0}(\tilde \s)$ appearing in the
potential take different forms for different choices of variables. For
instance, there might exist variables in terms of which the potential
is polynomial, whereas in another set it contains exponential
functions.

In the previous sections we expanded the Lagrangian around the
constant background \eqref{cs2}. The idea is that perturbations around
this background can be quantized and interpreted as particles. Their
tree-level masses and coupling constants are given by the coefficients
of the Taylor expansion around the point $\tilde\s=0$, i.e. by terms of the form
$\tilde{\mathcal K}_{\phi\phi}^{(n)}$ and $\tilde V^{(n)}$. Certainly, since the
functions depend on the variable choice, for different sets of variables, tree-masses and
coupling constants will take different values. Nevertheless, the equivalence
theorems of \cite{Coleman:1969sm} show that the so constructed quantum
theories are equivalent for \emph{all} choices of variables. A consequence
of  these theorems is that whenever one takes into account the whole
(possibly infinite) series of terms in the Lagrangian to compute
$S$-matrix elements, the result will not depend on the choice of
variables if the transformations are well-defined perturbatively. The
situations is different, however, if one uses effective field theory
arguments to truncate the Lagrangian because, as already mentioned,
the individual terms of the series expansions do depend on the choice
of variables.  This means that conditions like \eqref{deccond0},
\eqref{wc0} and \eqref{supcond0} depend on the choice of variables.
Therefore, applied to arbitrary variables, such conditions should be
considered as sufficient but \emph{not} necessary. It can happen, for
instance, that for some choice of variables some of the suppression
conditions \eqref{supcond0} do not hold, but that the corresponding
terms are nevertheless irrelevant\footnote{Technically, this can
happen in the following way. The interaction Lagrangian can contain
terms with big coefficients. These terms violate some of the
conditions \eqref{supcond0} and are therefore expected to be important
much below the scale $M$. However, there can be cancellations between
terms of the different series contained in \eqref{intlag0} which make
that also terms that violate the conditions \eqref{supcond0} can be
irrelevant.}. In order to have a variable independent statement, the
ensemble of conditions \eqref{deccond0}, \eqref{wc0} and
\eqref{supcond0} should be read as follows:

\noindent{\it``If there exists a set of variables in terms of
which the conditions \eqref{deccond0}, \eqref{wc0} and
\eqref{supcond0} hold, then, at energies well below $M$, the
scalar-field theory contained in  \eqref{fulltdscale} is
indistinguishable from the corresponding renormalizable theory.''}

\noindent Understood this way, the conditions are necessary
and sufficient.

As a particular example of the previous reasoning, 
one may wonder  whether there exists a set of field variables in terms
of which the kinetic part of the Lagrangian
(\ref{fulltdscale_diff_ein}) takes \emph{exactly} the canonical form.
The condition for such variables to exist is the vanishing of the
Riemann tensor computed from the field space metric
\cite{Cornwall:1974km}
\begin{equation}\label{metric}
\left\{{\mathcal K}_{ij}(\tilde{\s},\tilde{\phi})\right\}=
\begin{pmatrix}
\e_\s M^2&&0\\0&&\tilde {\mathcal K}_{\phi\phi}(\tilde{\s})
\end{pmatrix}\;.
\end{equation}
This condition corresponds to \footnote{In terms of the functions without tilde, the same condition
reads\\
$
{\mathcal K}_{\phi\phi}'(\s) \left({\mathcal K}_{\phi\phi}(\s)
 {\mathcal K}_{\s\s}'(\s)+{\mathcal K}'_{\phi\phi}(\s)
  {\mathcal K}_{\s\s}(\s)-2
{\mathcal K}_{\s\phi}(\s) {\mathcal K}_{\s\phi}'(\s)
\right)+2 \left( {\mathcal K}_{\s\phi}(\s)^2
-{\mathcal K}_{\phi\phi}(\s) {\mathcal K}_{\s\s}(\s)\right)
 {\mathcal K}_{\phi\phi}''(\s)=0\;.
$}
\begin{equation}\label{kincon}
\tilde{\mathcal K}_{\phi\phi}'(\tilde{\s})^2-2\tilde
{\mathcal K}_{\phi\phi}(\tilde{\s})\tilde{\mathcal K}_{\phi\phi}''
(\tilde{\s})=0\;.
\end{equation}
Functions $\tilde{\mathcal K}_{\phi\phi}(\tilde{\s})$ which satisfy this
equation have the form
\begin{equation}
\tilde{\mathcal K}_{\phi\phi}(\tilde{\s})=\left(c_1\,\tilde{\s}+c_2\right)^2\;,
\end{equation}
where $c_1$ and $c_2$ are arbitrary constants.
One can also formulate the conditions which guarantee that for the
variables that give a canonical kinetic term, the scalar field
potential (for $\Lambda_0=0$) becomes a polynomial of a maximum order $p$,
\begin{equation}
 \tilde{V}(\tilde{\s})_{;i_1;i_2;i_3;...;i_{p+1}}=0\;,
\end{equation}
where the semicolon stands for the covariant derivative with respect
to the metric \eqref{metric}. If these conditions hold for
$p=4$ and at the same time condition \eqref{kincon} is fulfilled, the scalar part of the Lagrangian describes a tree unitary and
renormalizable quantum field theory \cite{Cornwall:1974km}. For this
to be the case, the function $\tilde{V}(\tilde{\s})$ has to be of the
form
\begin{equation}\begin{array}{lcll}
\tilde V(\tilde \s)&=&v_0+v_1\,\tilde\s+v_2\,\tilde\s^2+v_3\,\tilde\s^3+v_4\,\tilde\s^4,&\quad\textrm{if}\quad c_1=0,\\
\tilde V(\tilde \s)&=&v_0+\tilde\s(\tilde\s+2c_2/c_1)\left(v_4\,\tilde\s^2+2c_2v_4/c_1\,\tilde\s+v_2-4c_2^2v_4/c_1^2\right),&\quad\textrm{if}\quad c_1\neq0,
\end{array}\end{equation}
where $v_0$, $v_1$, $v_2$, $v_3$ and $v_4$ are arbitrary constants.
If we also impose the conditions \eqref{maincond},
which correspond to
${\tilde{V}(0)=\tilde V'(0)=0}$,
we can further restrict the form of the function
$\tilde{V}(\tilde{\s})$ to
\begin{equation}\begin{array}{lcll}
\tilde V(\tilde \s)&=&v_2\,\tilde\s^2+v_3\,\tilde\s^3+v_4\,\tilde\s^4,&\quad\textrm{if}\quad c_1=0,\\
\tilde V(\tilde \s)&=&\tilde\s^2\left(v_2+v_4\,\tilde\s^2\right),&\quad\textrm{if}\quad c_1\neq0\;\;\textrm{and}\;\;c_2=0,\\
\tilde V(\tilde \s)&=&v_4\,\tilde\s^2\left(\tilde\s+2c_2/c_1\right)^2,&\quad\textrm{if}\quad c_1\neq0\;\;\textrm{and}\;\;c_2\neq0.\\
\end{array}\end{equation}

%%%%%%%%%%%%%%%%%%%%%%%%%%%%
\section{Including gauge bosons}
\label{ctb}
%%%%%%%%%%%%%%%%%%%%%%%%%%%%
In this section we will consider the addition of
(massive) gauge fields to the previous picture of scale-invariant TDiff Lagrangians. 
Remind that in the Higgs mechanism, gauge fields get their masses from a non-zero
expectation value of a scalar field. We are going to show how a
similar phenomenon can occur due to spontaneous breaking of scale
invariance in a scale-invariant TDiff theory, where the massive field $\s$
will play a role similar to the Higgs field of the SM. For simplicity
we will consider the case of an Abelian gauge group.

If the scalar field $\phi$  is promoted to a
complex field, the Lagrangian (\ref{fulltdscale}) is  invariant under a global $U(1)$
symmetry. This symmetry can be turned into a gauge symmetry by
introducing an Abelian gauge field (note that gauge fields have
scaling dimension $d_A=1$). The generalization of \eqref{fulltdscale} to this
case reads
\begin{equation}
\label{LU1}
\begin{split}
 \frac{\mathcal{L}}{\sqrt{-g}}=&-\frac{1}{2}|\phi|^2
f(-g)R-\frac{1}{2}|\phi|^2
{\mathcal G}_{gg}(-g)(\partial g)^2
-\frac{1}{2}{\mathcal G}_{\phi\phi}(-g)D\phi\cdot\left(D\phi\right)^*\\
&+\frac{1}{2}{\mathcal G}_{g\phi}^*(-g)\phi^* ~\partial g\cdot
D\phi+\frac{1}{2}{\mathcal G}_{g\phi}(-g)\phi ~\partial g\cdot \left(D
\phi\right)^*-\frac{1}{2}{\mathcal G}_{na}
(-g)~\partial|\phi|\cdot\partial|\phi|
\\
&-v(-g)|\phi\phi^*|^2-\frac{1}{4}{\mathcal G}_{AA}(-g)
F^2-\frac{1}{4}{\mathcal G}_\varepsilon(-g)
F\wedge F \;,
\end{split}
\end{equation}
where the covariant derivative is defined as $D_\mu\equiv
\partial_\mu-\mathrm{i}eA_\mu$ and the function ${\mathcal G}_{g\phi}(-g)$ is
complex valued. In this action we have also included the
non-analytical term $\pd |\phi|$. Notice that this term is \emph{unique} and perfectly well
defined around the background $\phi_0\neq 0$, so it is natural to consider
it as a term in the potential on the same footing as we consider generic TDF\footnote{Certainly, 
around the symmetry breaking background those terms involve non-renormalizable operators. Considering
the perturbation theory for certain TDF, those operators should be placed beyond the cut-off of the theory
which basically implies that the non-analytical term should be suppressed altogether. However, as we
emphasized in Section \ref{varchoice}, this conclusion depends on the 
choice of fields and certain higher order operators in one representation may be resummed to 
a renormalizable form by a local field redefinition.}. Moreover, we have defined the wedge
product as $F\wedge F=\e^{\m\n\r\s}F_{\mu\nu}F_{\r\s}$, where
$\e_{\m\n\r\s}\equiv\sqrt{-g}\,\varepsilon_{\m\n\r\s}$, with
$\varepsilon_{\m\n\r\s}$ being the standard Levi-Civita tensor.
We will analyze the theory in the unitary gauge $\phi^*=\phi$,
in which the Lagrangian reads
\begin{equation}
\begin{split}
 \frac{\mathcal{L}}{\sqrt{-g}}=&-\frac{1}{2}\phi^2
f(-g)R-\frac{1}{2}\phi^2
{\mathcal G}_{gg}(-g)(\partial
g)^2-\frac{1}{2}\left({\mathcal G}_{\phi\phi}(-g)+{\mathcal G}_{na}(-g)\right)
(\partial\phi)^2\\
&+\textrm{Re}\left[{\mathcal G}_{g\phi}(-g)\right]\phi
~\partial g\cdot
\partial\phi+e~\textrm{Im}\left[{\mathcal G}_{g\phi}(-g)\right]\phi^2
~\partial g\cdot A-\frac{1}{2}e^2~{\mathcal G}_{\phi\phi}(-g)
A^2\phi^2\\&-v(-g)\phi^4-\frac{1}{4}{\mathcal G}_{AA}(-g)
F^2-\frac{1}{4}{\mathcal G}_{\varepsilon}(-g)
F\wedge F\;,
\end{split}
\end{equation}
where $\textrm{Re}$ and $\textrm{Im}$ stand for the real and 
imaginary part, respectively. Following the formalism developed in
Section \ref{sec_Diff}, one can directly write down the equivalent Diff
invariant theory in the Einstein frame as (see (\ref{eq:Eins_scale}))
\begin{equation}\label{Egauge}
\begin{split}
 \frac{\mathcal{L}_e}{\sqrt{-\tilde g}}=&-\frac{1}{2}M^2
\tilde R-\frac{1}{2}M^2
{\mathcal K}_{\s\s}(\s)(\partial\s)^2-\frac{1}{2}M^2{\mathcal K}_{\phi\phi}
(\s)(\partial\ln(\phi/M))^2\\
&-M^2{\mathcal K}_{\s\phi}(\s)~\partial\s\cdot
\partial\ln(\phi/M)-e M^2{\mathcal K}_{\s A}(\s)~
\partial \s\cdot
A-\frac{1}{2}e^2M^2{\mathcal K}_{int}(\s)
A^2\\&-M^4V(\s)-\frac{1}{4} {\mathcal K}_{AA}(\s)
F^2-\frac{1}{4}{\mathcal K}_\varepsilon(\s)
F\wedge F -\frac{M^4\Lambda_0}{\phi^4f(\s)^2\sqrt{\s}}\;,
\end{split}
\end{equation}
where
\begin{equation}\label{Ks}\begin{array}{lcllcl}
{\mathcal K}_{\s\s}(\s)&=&\dfrac{{\mathcal G}_{gg}(\s)}{f(\s)}
+\dfrac{3}{2}\left(\dfrac{f'(\s)}{
f(\s)} \right)^2,&\quad
{\mathcal K}_{\phi\phi}(\s)&=&\dfrac{{\mathcal G}_{\phi\phi}(\s)
+{\mathcal G}_{na}(\s)}{f(\s)}+6,\\
{\mathcal K}_{\s\phi}(\s)&=&\dfrac{\textrm{Re}
\left[{\mathcal G}_{g\phi}(\s)\right]}{f(\s)}
+3\dfrac{f'(\s)}{f(\s)},&\quad {\mathcal K}_{\s A}(\s)&=&\dfrac{\textrm{Im}\left[{\mathcal G}_{g\phi}(\s)\right]}{f(\s)},\\ 
{\mathcal K}_{int}(\s)&=
&\dfrac{{\mathcal G}_{\phi\phi}(\s)}{f(\s)}
,&\quad V(\s)&=&\dfrac{v(\s)}{f(\s)^2},\\
{\mathcal K}_{AA}(\s)&=&{\mathcal G}_{AA}(\s),&\quad {\mathcal K}_{\varepsilon}(\s)&=&{\mathcal G}_{\varepsilon}(\s).
\end{array}\end{equation}
At this point, as in the case without gauge fields, we can make a
field redefinition in order to eliminate the derivative couplings
between the different fields. This will simplify the interpretation of
the theory as a description of interacting particles. The extension of expression
\eqref{tr0} to this case is \footnote{We assume that ${\mathcal K}_{\phi\phi}$, ${\mathcal K}_{int}$
and ${\frac{{\mathcal K}_{\s\s}{\mathcal K}_{\phi\phi}-{\mathcal
K}_{\s\phi}^2}{{\mathcal K}_{\phi\phi}} -\frac{{\mathcal K}_{\s
A}^2}{{\mathcal K}_{int}}}$ are non-vanishing.}
\begin{gather}
 \tilde{\sigma}=\int_{\sigma_0}^\sigma \di\sigma'
\sqrt{\left|\frac{{\mathcal K}_{\s\s}{\mathcal K}_{\phi\phi}
-{\mathcal K}_{\s\phi}^2}{{\mathcal K}_{\phi\phi}}
-\frac{{\mathcal K}_{\s A}^2}{{\mathcal K}_{int}}\right|}
\;,
\quad \tilde{\phi}=M\left(\ln\frac{\phi}{M}+
\int_{\sigma_0}^\sigma \di\sigma'
\frac{{\mathcal K}_{\s\phi}}{{\mathcal K}_{\phi\phi}}\right)\;,\nonumber\\
\tilde{A}_\mu=A_\mu+\frac{1}{e}
\frac{{\mathcal K}_{\s A}}{{\mathcal K}_{int}}\partial_\mu\s\;,\label{trgauge}
\end{gather}
in terms of which the above Lagrangian reads
\begin{equation}
\label{eq:Lagrgauge}
 \begin{split}
\frac{\mathcal{L}_e}{\sqrt{-\tilde g}} 
=&-\frac{1}{2}M^2\tilde{R}-\frac{1}{2}\e_\s M^2
(\partial\tilde\sigma)^2
-\frac{1}{2}\tilde{\mathcal K}_{\phi\phi}(\tilde{\sigma})
(\partial\tilde{\phi})^2\\
&-\frac{1}{2}e^2\tilde{\mathcal K}_{int}(\tilde{\sigma})M^2\tilde{A}^2
-\frac{1}
{ 4 } \tilde{\mathcal K}_{AA}(\tilde{\sigma})
\tilde{F}^2-\frac{1}{4}\tilde{\mathcal K}_{\varepsilon}(\tilde{\sigma})
\tilde{F}\wedge\tilde{F}\\
&-\tilde{V}(\tilde{\sigma})M^4
-\Lambda_0\, \tilde{\mathcal K}_{\Lambda_0}\exp\left(-\frac{4\tilde\phi}{M}\right),
\end{split}\;
\end{equation}
where
$\e_\s=\textnormal{sign}\left(\frac{{\mathcal K}_{\s\s}{\mathcal
K}_{\phi\phi}-{\mathcal K}_{\s\phi}^2} {{\mathcal K}_{\phi\phi}}
-\frac{{\mathcal K}_{\s A}^2}{{\mathcal K}_{int}}\right)$ and
$\tilde{\mathcal K}_{\Lambda_0}(\tilde \s)$ is defined in \eqref{eq:tildepot}.
Note that the field $\tilde\phi$ is completely decoupled from the vector fields
(which follows from scale and gauge invariance), thus the mass of the
vector bosons is related to the interaction with
the ``gravitational'' field $\tilde\sigma$. In this loose sense, the role of
the Higgs field is played by the determinant of the metric\footnote{A possible connection between the Higgs field and the determinant of the metric was suggested previously in \cite{Chernodub:2008rz,Faddeev:2008qc} from different considerations.}.
 The previous Lagrangian may be subject
to different constraints that we will consider in the next sections.

%%%%%%%%%%%%%%%%%%%%%%%%%%%%
\subsection{Local degrees of freedom}
%%%%%%%%%%%%%%%%%%%%%%%%%%%%
Like in the case without gauge fields, the existence of a constant
solution $\tilde{g}_\mn=\eta_\mn$, $\tilde{\s}=0$,
$\tilde{\phi}=\tilde{\phi}_0$ and $\tilde{A}_\mu=0$ is assured by the
conditions (we also assume $f(\s_0)\neq 0$)
\begin{equation}
\label{maincond1}
v(\s_0)=v'(\s_0)=0\;.
\end{equation} 
Let us also recall that the constant solution has $\Lambda_0=0$. We again
want to examine the nature of the perturbations around the constant
solution, which we define as
\begin{equation}\begin{array}{lcllcl}
 \tilde{g}_\mn&=&\eta_\mn+\dfrac{\tilde{h}_\mn}{M}\;,&\quad \tilde{\s}
 &=&\dfrac{\varsigma}{M}\;,\\
\tilde{\phi}&=&\tilde{\phi}_0+\dfrac{\varphi}{\left|\tilde{
\mathcal K}_{\phi\phi}^{(0)}\right|^{1/2}}\;,&\quad
\tilde{A}_\mu&=&\dfrac{\tilde{A}_\mu^c}
{\left|\tilde{\mathcal K}_{AA}^{(0)}\right|^{1/2}} \;.
\end{array}\end{equation}
\vspace{0.2cm}

\noindent In the rest of Section \ref{ctb}, Lorentz indices are raised, lowered
and contracted with the Minkowski metric $\eta_{\m\n}$.
To quadratic order the Lagrangian (\ref{eq:Lagrgauge}) reduces to 
\vspace{0.4cm}
\begin{align}
 \mathcal{L}_e^Q=&\tilde{\mathcal{L}}_{EH}^Q
-\e_\varsigma\frac{1}{2}(\partial\varsigma)^2
-\e_\varphi\frac{1}{2}(\partial\varphi)^2
-\frac{1}{2}m_\varsigma^2\varsigma^2
-\e_A\frac{1}{4}\tilde{F}^2-\frac{1}{2}m_A^2\tilde{A}^2\; ,
\end{align} 
where
\begin{align}
&\epsilon_\varsigma\equiv
\textnormal{sign}\left(\frac{{\mathcal K}_{\s\s}^{(0)}
{\mathcal K}_{\phi\phi}^{(0)}-\left({\mathcal K}_
{ \s\phi}^{(0)}\right)^2}{{\mathcal K}_{ \phi\phi}^{(0)}}
-\frac{\left({\mathcal K}_{\s A}^{(0)}\right)^2}
{{\mathcal K}_{int}^{(0)}}\right)\;,
\quad\epsilon_\varphi\equiv \textnormal{sign}\left({\mathcal K}_{
\phi\phi}^{(0)}\right)\;,\quad
\epsilon_A\equiv \textnormal{sign}\left({\mathcal K}_{AA}^{(0)}\right)\;,\nonumber\\
&\quad\hspace{3cm} \quad \quad \quad \quad \quad \quad \quad m_A^2\equiv \e_A e^2\frac{{\mathcal K}_{int}^{(0)}}
{{\mathcal K}_{A A}^{(0)}}M^2\;,\nonumber\\
&\quad \quad  \quad  m_\varsigma^2\equiv \e_\varsigma\tilde{V}^{(2)}M^2
=\e_\varsigma\left(\frac{{\mathcal K}_{\s\s}^{(0)}
{\mathcal K}_{\phi\phi}^{(0)}-\left({\mathcal K}_
{ \s\phi}^{(0)}\right)^2}{{\mathcal K}_{ \phi\phi}^{(0)}}
-\frac{\left({\mathcal K}_{\s A}^{(0)}\right)^2}{{\mathcal K}_{int}^{(0)}}
\right)^{-1}V^{(2)}M^2.
\label{eq:mass_g}
\end{align}
At the level of the quadratic Lagrangian, the following conditions must be
satisfied:
\begin{itemize}
 \item For positive definite kinetic terms (absence of ghosts):
\be\label{kc1} 
\epsilon_\varsigma,\;\epsilon_\varphi,\;\epsilon_A=1\;.
\ee
\item For positive or zero masses (absence of
tachyons):
\be\label{mc1} 
m_\varsigma^2,\;m_A^2\geq 0\;.
\ee
\end{itemize}

%%%%%%%%%%%%%%%%%%%%%%%%%%%%
\subsection{Interactions and separation of scales}
%%%%%%%%%%%%%%%%%%%%%%%%%%%%
The terms of dimension up to four are
\begin{equation}\begin{split}
 \mathcal{L}_e^{\textnormal{int}\leq
4}=&-\frac{1}{3!}\k_\varsigma
\varsigma^3-\frac{\l_\varsigma}{4!}\varsigma^4-\frac{1}{2}\k_A\eta^{
\m\n}
\tilde {A}^c_\m\tilde{A}^c_\n\varsigma-\frac{1}{4}\l_A\eta^{\m\n}
\tilde {A}^c_\m\tilde{A}^c_\n\varsigma^2\\&-\frac{1}{4}\frac{
m_\varsigma^2 } { M }
\varsigma^2 \tilde{h}-\frac{1}{16}\frac{m_\varsigma^2}{M^2}
\varsigma^2\left(\tilde{h}^2-2\tilde{h}_{\m\n}\tilde{h}^{\m\n}
\right)-\frac{1}{12}\frac{ \k_\varsigma}{M}
\varsigma^3\tilde{h}
\\
&-\frac{1}{4}\frac{m_A^2}{M}
\tilde {A}^c_\m\tilde{A}^c_\n
\left(\eta^{\m\n}\tilde{h}-2\tilde h^{\m\n}\right)-\frac{1}{4}\frac{\k_A}{M}
\varsigma \tilde A_\m^c \tilde A_\n^c
\left(\eta^{\m\n}\tilde h-2\tilde h^{\m\n}\right)\\
&-\frac{1}{16}\frac{m_A^2}{
M^2 }
\tilde{A}^c_\m\tilde{A}^c_\n\left(\eta^{\m\n}\tilde{h}^2
-4\tilde h^{\m\n}\tilde h-2\eta^{\m\n}\tilde{h}_{\r\s}\tilde{h}^{
\r\s}
+8\tilde h^\m_\r \tilde h^{\r\n}\right)\;,
\end{split}\end{equation} 
where we have defined the parameters
\begin{align}
&\k_\varsigma\equiv\tilde{V}^{(3)}M\;,\quad \l_\varsigma
\equiv\tilde{V}^{(4)}\;,
\quad\k_A\equiv e^2\frac{\tilde{\mathcal K}_{int}^{(1)}}
{\tilde {\mathcal K}_{AA}^{(0)}}M\;,
\quad\l_A\equiv e^2\frac{\tilde{\mathcal K}_{int}^{(2)}}
{\tilde {\mathcal K}_{AA}^{(0)}}\;.
\end{align}
As in the previous section, we
will require that the mass scales of the
fields $\varsigma$ and $\tilde A_\m$ 
are parametrically smaller than the cut-off $M$,
\begin{align}
 &\left|\frac{m_\varsigma}{M}\right|=\sqrt{\left|\tilde{V}^{(2)}
\right|}\ll 
1\;, \quad 
\left|\frac{m_A}{M}\right|=
\sqrt{\left|\tilde e^2 {\mathcal K}_{int}^{(0)}
\right|}\ll 1\;,
\label{deccond1}
\end{align}
with the definition $\tilde e^2\equiv\frac{e^2}{{\mathcal
K}_{AA}^{(0)}}$. In addition, we have the following conditions that
prevent the theory from being strongly coupled,
\begin{align}
\label{eq:decop_gauge}
 &\left\{\frac{\kappa_\varsigma}{m_{min}},\frac{\kappa_A}{m_{min}},\l_\varsigma,\l_A,\right\}
  \lesssim 1\;,
 \end{align}
where $m_{min}\equiv \min{(m_{\varsigma},m_A)}$. 
Note that the first two conditions might not be necessary for a particular structure 
of the interactions. In particular, these conditions are not necessary if the theory corresponds to the Abelian Higgs model.

The higher dimensional terms can be written schematically as
\begin{align}
\mathcal{L}_e^{\textnormal{int}>4}=\phantom{+}&\sum_{ n_h>0 }
^\infty\frac{1}{M^{n_h}}\left(\mathcal{L}_e^Q+\mathcal{L}_e^{
\textnormal{int}\leq 4}\right)\tilde{h}^{n_h}\nonumber\\
+&\sum_{\substack {n_h\geq 0\\
n_\varsigma>0}}
^\infty\left(\frac{1}{M_{\phi\phi}(n_h,n_\varsigma)}\right)^{
n_h+n_\varsigma }
(\partial\varphi)^2
\tilde{h}^{n_h}\varsigma^{n_\varsigma}+\sum_{\substack{n_h\geq
0\\n_\varsigma>4}}
^\infty\left(\frac{1}{M_V(n_h,n_\varsigma)}\right)^{n_h+n_\varsigma-4}
\tilde{h}^ {n_h} \varsigma^{ n_\varsigma}\nonumber\\
+&\sum_{\substack{n_h\geq 0\\n_\varsigma>2}}
^\infty\left(\frac{1}{M_{int}(n_h,h_\varsigma)}\right)^{n_h
+n_\varsigma-2}
(\tilde A^c)^2 \tilde h^{n_h}\varsigma^{n_\varsigma}\nonumber\\
+&\left(\sum_{\substack{n_h\geq 0\\n_\varsigma>0}}
^\infty\left(\frac{1}{M_{AA}(n_h,n_\varsigma)}\right)^{
n_h+n_\varsigma }
+\sum_{\substack{n_h\geq
0\\n_\varsigma>0}}
^\infty\left(\frac{1}{M_\varepsilon(n_h,n_\varsigma)}\right)^{
n_h+n_\varsigma }
\right)\partial^2(\tilde A^c)^2 \tilde h^{n_h}\varsigma^{n_\varsigma}
\;,\label{intlag1}
\end{align}
where, as before, we neglect numerical factors of order one, neglect
tensor  indices and define the suppression scales
\begin{equation}\begin{array}{lcllcl}
M_{\phi\phi}(n_h,n_\varsigma)&\sim &M\left|\dfrac
{\tilde {\mathcal K}_{\phi\phi}^{(n_\varsigma)}}{\tilde
{\mathcal K}_{\phi\phi}^{(0)}}\right|^{\frac{-1}{n_h+n_\varsigma}},
 &\quad M_V(n_h,n_\varsigma)&\sim &M\left|\tilde V^{
(n_\varsigma)}\right|^{\frac{-1}{n_h+n_\varsigma-4}},\\
 M_{int}(n_h,n_\varsigma)&\sim &M\left|\tilde e^2~
\tilde {\mathcal K}_{int}^{(n_\varsigma)}\right|^{\frac{-1}{
n_h+n_\varsigma}},&\quad M_{AA}(n_h,n_\varsigma)&\sim &M\left|
\dfrac{\tilde {\mathcal K}_{AA}^{
(n_\varsigma)}}{\tilde {\mathcal K}_{AA}
^{(0)}}\right|^{\frac{-1}{n_h+n_\varsigma}},\\
 M_\varepsilon(n_h,n_\varsigma)&\sim &M\left|
\dfrac{\tilde {\mathcal K}_\varepsilon^{(n_\varsigma)}}
{\tilde {\mathcal K}_{AA}
^{(0)}}\right|^{\frac{-1}{n_h+n_\varsigma}}.
\end{array}\end{equation}
The first term in \eqref{intlag1} represents the standard higher 
dimensional operators of a theory minimally coupled to gravity, and
are suppressed at energies below $M$ as soon as the conditions \eqref{deccond1} and (\ref{eq:decop_gauge}) hold. The additional operators come
with the suppression scales $M_{\phi\phi}(n_h,n_\varsigma)$,
$M_V(n_h,n_\varsigma)$, $M_{int}(n_h,n_\varsigma)$,
$M_{AA}(n_h,n_\varsigma)$ and $M_\varepsilon(n_h,n_\varsigma)$. These
are comparable to or bigger than the scale $M$ whenever
\begin{equation}
\label{supcond1}
\begin{array}{lcllcllcl}
\left|\dfrac
{\tilde {\mathcal K}_{\phi\phi}^{(n_\varsigma)}}{\tilde
{\mathcal K}_{\phi\phi}^{(0)}}\right|^{\frac{1}{n_h+n_\varsigma}}&\leq &1\;,
&\quad\left|\tilde V^{
(n_\varsigma)}\right|^{\frac{1}{n_h+n_\varsigma-4}}&\leq &1\;,
&\quad\left|\tilde e^2
\tilde {\mathcal K}_{int}^{(n_\varsigma)}\right|^{\frac{1}{n_h+n_\varsigma}}
&\leq &1\;,\\
\left|\dfrac{\tilde {\mathcal K}_{AA}^{
(n_\varsigma)}}{\tilde {\mathcal K}_{AA}
^{(0)}}\right|^{\frac{1}{n_h+n_\varsigma}}&\leq &1\;,&\quad\left|
\dfrac{\tilde {\mathcal K}_\varepsilon^{(n_\varsigma)}}{\tilde {\mathcal K}_{AA}
^{(0)}}\right|^{\frac{1}{n_h+n_\varsigma}}&\leq &1\;,
\end{array}
\end{equation}
for all values $n_\varsigma$ and $n_h$ can take in the sums in
\eqref{intlag1}. If the conditions \eqref{kc1}, \eqref{mc1},
\eqref{deccond1}, \eqref{eq:decop_gauge}, (\ref{eq:decop_gauge}) and \eqref{supcond1} are met,
the effective Lagrangian for describing the scalar and vector sectors
at energies far below $M$ is
\begin{equation}
\begin{split}
 \mathcal{L}_e^{\textnormal{eff}}=
&-\frac{1}{2}(\partial\varsigma)^2
-\frac{1}{2}(\partial\varphi)^2
-\frac{1}{2}m_\varsigma^2\varsigma^2-\frac{1}{3!}
\k_\varsigma\varsigma^3-\frac{\l_\varsigma}{4!}
\varsigma^4\\
&-\frac{1}{4}(\tilde{F}^c)^2
-\frac{1}{2}m_A^2(\tilde{A}^c)^2
-\frac {1}{2}
\k_A(\tilde {A}^c)^2\varsigma-\frac{1}{4}\l_A
(\tilde{A}^c)^2\varsigma^2\;.
\end{split}
\end{equation} 
We would like this Lagrangian to give rise to a consistent quantum
field theory at energies low with respect to $M$. It has been shown
\cite{Cornwall:1974km} that the only tree-unitary theories containing
scalar fields and massive vector particles are those that correspond
to a spontaneously broken gauge theory\footnote{For theories with a
conserved current, like  in fermionic theories, the models where Abelian
massive fields interact just with the conserved current are also
allowed \cite{Cornwall:1974km} .}. Thus, for our model to be
tree-unitary at energies below $M$ (and above $m_A$), the above effective Lagrangian
should correspond to the Abelian Higgs model in the unitary gauge.
This means that the six couplings $m_\varsigma$, $\k_\varsigma$, $\l_\varsigma$,
$m_A$, $\k_A$ and $\l_A$ should satisfy the three relations
\begin{align}
 &\frac{\l_\varsigma}{\l_A}=\frac{\k_\varsigma}{\k_A}, 
 \quad \frac{\l_\varsigma}{\l_A}=\frac{3}{2}\frac{m_\varsigma^2}{m_A^2},
 \quad m_\varsigma^2=\frac{1}{3}\frac{\k_\varsigma^2}{\l_\varsigma}\;.
\end{align}
In the present model, these relations can be translated to
the following conditions on the TDF:
\begin{align}
&\frac{\tilde {\mathcal K}_{int}^{(0)}}{\tilde {\mathcal K}_{int}^{(1)}}
\simeq\frac{1}{2}
\frac{\tilde {\mathcal K}_{int} ^{(1)}}{\tilde {\mathcal K}_{int}^{(2)}}\;,
\quad
\frac{\tilde V^{(2)}}{\tilde V^{(3)}}\simeq\frac{2}{3}
\frac{\tilde {\mathcal K}_{int}^{(0)}}{\tilde {\mathcal K}_{int}^{(1)}}\;,
\quad
\frac{\tilde V^{(2)}}{\tilde V^{(3)}}\simeq
\frac{1}{3}\frac{\tilde V^{(3)}}{\tilde V^{(4)}}
\label{abhiggscond}\;,
\end{align}
where by the approximate equalities we mean that the relation should
hold up to suppressed terms, i.e. for two quantities $a$ and $b$ one has
$a\simeq b$ whenever $a=b\left(1+\mathcal{O}\left(
\frac{m_\varsigma}{M},\frac{\kappa_\varsigma}{M},\frac{m_A}{M},
\frac{\kappa_A}{M}\right)\right)$. These conditions are expected to be
stable under radiative corrections since they approximately correspond to a
 gauge theory with spontaneous symmetry breaking. 

We can now draw the following conclusion. If there exists a set of
variables in terms of which the conditions \eqref{maincond1},
\eqref{kc1}, \eqref{mc1}, \eqref{deccond1}, \eqref{eq:decop_gauge},
\eqref{supcond1} and  \eqref{abhiggscond} hold, then at energies
well below $M$ the theory given by \eqref{LU1} is indistinguishable
from the renormalizable Abelian Higgs model. While some of these conditions
can be naturally satisfied, for instance by polynomial TDF, the conditions
related to the smallness of particle masses with respect to
the Planck scale $M$ may require a fine-tuning (see Section
\ref{potentials}). Hence, scale-invariant TDiff theories do not provide
any explanation for the huge difference between the Planck mass $M$ and
the mass scales of the SM. However, the presence of the extra field $\phi$ allows
to introduce scale-invariant regularization schemes, under which the mass of the
Higgs boson is not affected by the cut-off scale of the theory \cite{Shaposhnikov:2008xi}.

%%%%%%%%%%%%%%%%%%%%%%%%
\section{Coupling to fermionic matter}
\label{ctm}
%%%%%%%%%%%%%%%%%%%%%%%%

Finally, let us study the inclusion of fermions to scale-invariant TDiff theories
\footnote{For the sake of illustration we only consider Dirac
spinors. Still the conclusions are generic as they only depend on the
dimensionality of the fields. In this context see also \cite{Blas:2008ce,Blas:2008uz}
for the first order formalism of unimodular gravity.}.
A generic scale-invariant spinor Lagrangian compatible with TDiff can be
written as\footnote{In this section we use the conventions of \cite{Fujii:2003pa}.}
\be\label{spinL}
{\mathcal L}_\psi=-b\,{\mathcal G}_\psi(b^2)\,\bar \psi\, b^{\m a}  \gamma_a
\left(\pd_\m+\frac{1}{8}[\gamma_c,\gamma_d] 
\,\omega^{\phantom{\m}cd}_\m\right)
\psi-b\,\phi\, {v}_\psi(b^2) \bar \psi \,\psi,
\ee 
where $b^{\m a}$ represents the inverse vierbein related to the metric
through $g_{\m\n}=\eta_{ac}\,b_\m^{\phantom{\m}a}b_\n^{\phantom{\n}c}\,,$
$\omega^{\phantom{\m}cd}_\m$ is its spin connection (see e.g. \cite{Fujii:2003pa}) and
$b=det[b_\m^{\phantom{\m}a}]=\sqrt{-g}$. Note that the fermionic  fields have scaling dimension $d_\psi = 3/2$.
This is the most general Lagrangian if one requires polynomiality in the fields $\phi$ and $\psi$.

We should mention here that, as soon as a theory includes several fields with non-trivial scaling dimensions, scale invariance alone does not forbid the presence of arbitrary functions of scale-invariant field combinations. In the present example,  all terms in the Lagrangian can in principle contain arbitrary functions of the combination $\bar\psi\psi/\phi^3$. Terms with $\phi$ in the denominator would be well-defined in a perturbative theory around a symmetry breaking background $\phi_0\neq 0$, however, they would correspond to higher dimensional operators. Terms with $\bar\psi\psi$ in the denominator, on the other hand, are in general ill-defined. We will stick to the requirement of polynomiality, bearing in mind that this is a variable dependent criterium.

Introducing the St\"uckelberg field as described in Section
\ref{sec_Diff}, the Lagrangian \eqref{spinL} can be written as
\be\label{fermions}
{\mathcal L}_\psi=-b\, {\mathcal G}_\psi(\s)\,\bar \psi\, b^{\m a}  \gamma_a
\left(\pd_\m+\frac{1}{8}[\gamma_c,\gamma_d]
\,\omega^{\phantom{\m}cd}_\m\right)
\psi-b\,\phi\, {v}_\psi(\s) \bar \psi \,\psi.
\ee 
In the vierbein formalism the field redefinition \eqref{eq:Eins_scale} corresponds to ${\tilde b_\m^{\phantom{\m}a}=\Omega\, b_\m^{\phantom{\m}a}}$. Together with the redefinition of the spinor field
$$
\tilde\psi=\Omega^{-3/2}\, \psi,
$$
it yields the Lagrangian in the E-frame (see e.g. \cite{Fujii:2003pa})
\be
{\mathcal L}_\psi= - \tilde b\,{\mathcal G}_\psi(\s)\, \bar{\tilde \psi}\,\tilde b^{\m i}
\gamma_i \left(\pd_\m+\frac{1}{8}[\gamma_j,\gamma_k]
\,\tilde\omega^{\phantom{\m}jk}_\m\right)
\tilde\psi-\tilde b\,\frac{M {v}_\psi(\s)}{\sqrt{f(\s)}} 
\bar{\tilde \psi}
\,\tilde\psi.
\ee
We see that the scale invariance of the spinor Lagrangian in the J-frame also leads to the decoupling of fermions from the dilaton field $\phi$ in the E-frame.

The above Lagrangian contains the interactions between the fundamental fermions, the gravitational field and the scalar field.
To study non-relativistic processes, it is convenient to formulate these interactions in terms of particles interacting through certain potentials.
For fields without strong interactions at low energies, this is done by a WKB approximation, and realizing that the corresponding particles 
propagate in geodesics of the metric to which they are coupled \cite{Alvarez:2008zw} (alternatively, one may use non-relativistic
scattering amplitudes and the Born approximation to reconstruct the potential characterizing the interaction \cite{BjerrumBohr:2004mz}). For other fields (such as quarks), the
consequence of non-trivial couplings in low-energy phenomenology (e.g. for the gravitational interaction of hadrons) is 
certainly more complicated \cite{Damour:2010rp,Kaplan:2000hh,Hui:2010dn}. In the present case, it is relatively 
simple to write down the different possible terms that can appear for the point-particle Lagrangians. They will be of the form
\be
{\mathcal L}_{pp}=\int \di \tau\sqrt{\phi^2\, {\mathcal G}_{pp}(\s)g_{\m\n}\frac{\di x^\m}{\di \tau}\frac{\di x^\n}{\di \tau}},
\ee
where $x^\m(\tau)$ denotes the worldline of a point particle and ${\mathcal G}_{pp}(\s)$ is an arbitrary function to be deduced from (\ref{fermions}). As happens for the fundamental 
fields, moving to the E-frame makes the field $\phi$ disappear ($\phi$  is not coupled to matter fields), and we are back to
a theory where particles move on geodesics of the effective metric ${\mathcal G}_{pp}(\s)\tilde g_{\m\n}$, reflecting
the fact that the fundamental fields are coupled to the fields $\tilde g_{\m\n}$ and $\s$. The interaction mediated by $g_{\m\n}$ is long-ranged, while the range of the interaction due to $\s$ depends on its mass $m_\varsigma$ (cf. \eqref{eq:mass_g}).

%%%%%%%%%%%%%%%%%%%%%%%%%%%%%%%%%%%%%%%%%%%%%%
\section{Application to the Standard Model}
\label{asm}
%%%%%%%%%%%%%%%%%%%%%%%%%%%%%%%%%%%%%%%%%%%%%%

The basics established in the preceding sections can be used to
construct a scale-invariant version of the Standard Model of particle
physics coupled to gravity. Let us describe how this should be done.
The scalar-tensor sector of the theory is given by the Lagrangian
\eqref{fulltdscale} where $\phi$ is replaced by the complex
Higgs-doublet $H$. All fermions and bosons of the SM are then added
and coupled to gravity in the way described in Sections \ref{ctb}
and \ref{ctm}, again with $H$ replacing $\phi$. The generalization
to the group structure of the SM is straightforward. All
TDF have to be chosen such that they fulfill a
series of conditions of the type of \eqref{maincond}, \eqref{kc1},
\eqref{mc1}, \eqref{deccond1}, \eqref{eq:decop_gauge}, \eqref{supcond1} and 
\eqref{abhiggscond}. In this way, one obtains a model whose particle
phenomenology at energies well below the Planck mass $M$ is
indistinguishable from that of the SM. In particular, the massless
dilaton practically decouples from all the fields of the SM, except for the Higgs field to which 
it couples only through very suppressed interactions.

%%%%%%%%%%%%%%%%%%%%%%%%%%%%
\section{Particular choices of the theory-defining functions}
\label{potentials}
%%%%%%%%%%%%%%%%%%%%%%%%%%%%
In the previous sections we have derived a number of conditions to be
satisfied by the theory-defining functions. These conditions are
summarized in Table \ref{table}.  Similar conditions  should be
imposed for the fermionic sector, but for the sake of simplicity we
will restrict our considerations to the scalar and gauge sector (Lagrangian (\ref{LU1})).
\renewcommand\arraystretch{1.3}
\begin{table}[h,c]\begin{center}
\begin{tabular}[]{|c|l|c|}\hline
&\multicolumn{1}{c|}{\textit{Physical Meaning}} & \textit{Formal
Conditions}\\\hline
C1&\begin{tabular}{ll}Existence of a constant flat solution\end{tabular} & $v(\s_0)=v'(\s_0)=0$\\\hline
C2& \begin{tabular}{ll}Induced gravitational coupling\end{tabular}& $f(\s_0)\neq 0$\\\hline
C3&\begin{tabular}{ll}Positive definite kinetic terms \\
(absence of ghosts)\end{tabular} &
$\e_\varsigma$, $\e_\varphi$, $\e_A$ $=1$\\\hline
C4&\begin{tabular}{ll}No negative masses \\ (absence of tachyons)\end{tabular}
&$m_\varsigma^2$, $m_A^2$ $\geq 0$\\\hline
C5&\begin{tabular}{ll}Decoupling of gravitational interactions\end{tabular} &
 $m_\varsigma$, $m_A$ $\ll M$\\\hline
C6&\begin{tabular}{ll}No strong coupling \end{tabular}& \begin{tabular}{cc}$\k_\varsigma$, $\k_A \lesssim \min(m_\varsigma,m_A)$\\
$\l_\varsigma$, $\l_A$ $\lesssim 1$\end{tabular}\\\hline
C7&\begin{tabular}{ll}Suppression of higher-dimensional operators \end{tabular}& $M_{\phi\phi}$, $M_V$,
$M_{int}$, $M_{AA}$, $M_\varepsilon$ $\gtrsim M$\\\hline
C8&\begin{tabular}{ll}Equivalence with Abelian Higgs model \end{tabular}&
$\dfrac{\k_A}{\l_A}\simeq\dfrac{\k_\varsigma}
{\l_\varsigma}\simeq 3\dfrac{m_\varsigma^2}
{\k_\varsigma}\simeq 2\dfrac{m_A^2}{\k_A}$\\\hline
\end{tabular}\end{center}
\caption{Conditions to be imposed on the theory-defining functions (TDF)}
\label{table}
\end{table}

The parameters in terms of which the conditions are formulated are
defined through the TDF. They are summarized in
Table 2 \Big(remember that $\tilde e^2\equiv\frac{e^2}{{\mathcal
L}_{AA}^{(0)}}$\Big).
\begin{table}[h,c]\begin{center}
\begin{tabular}[]{|c|l|c|}\hline
i.& Signs of kinetic terms&
\begin{tabular}{l}
$\e_\varsigma=\textnormal{sign}\left({\mathcal K}_{\s\s}^{(0)}
-\frac{\left({\mathcal K}_
{\s\phi}^{(0)}\right)^2}{{\mathcal K}_{\phi\phi}^{(0)}}
-\frac{\left({\mathcal K}_{\s A}^{(0)}\right)^2}
{{\mathcal K}_{int}^{(0)}}\right)$\\
$\epsilon_\varphi\equiv \textnormal{sign}\left({\mathcal K}_{
\phi\phi}^{(0)}\right)$\\
$\epsilon_A\equiv \textnormal{sign}\left({\mathcal K}_{AA}^{(0)}\right)$
\end{tabular}\\\hline
ii.&  \begin{tabular}{ll}Masses and\\  relevant couplings
\end{tabular} &
 \begin{tabular}{lll}
$m_\varsigma^2\equiv\e_\varsigma\tilde{V}^{(2)}M^2$,
&$\k_\varsigma\equiv\tilde{V}^{(3)}M$,
&$\l_\varsigma\equiv\tilde{V}^{(4)}$,
\\
$m_A^2\equiv \e_A \tilde e^2 {\mathcal K}_{int}^{(0)}M^2$,
&$\k_A\equiv \tilde e^2\tilde{\mathcal K}_{int}^{(1)}M$,
&$\l_A\equiv \tilde e^2\tilde{\mathcal K}_{int}^{(2)}$,
 \end{tabular}
\\\hline
iii.&Suppression scales&
 \begin{tabular}{lll}
\\  $M_{\phi\phi}(n_h,n_\varsigma)\sim M\left|\frac
{\tilde {\mathcal K}_{\phi\phi}^{(n_\varsigma)}}{\tilde
{\mathcal K}_{\phi\phi}^{(0)}}\right|^{\frac{-1}{n_h+n_\varsigma}},$
&$n_h\geq0, \;n_\varsigma>0$\\
$M_V(n_h,n_\varsigma)\sim M\left|\tilde V^{
(n_\varsigma)}\right|^{\frac{-1}{n_h+n_\varsigma-4}},$
&$n_h\geq0, \;n_\varsigma>4$\\
$M_{int}(n_h,n_\varsigma)\sim M\left|\tilde e^2~
\tilde {\mathcal K}_{int}^{(n_\varsigma)}\right|^{\frac{-1}{
n_h+n_\varsigma}},$
&$n_h\geq0, \;n_\varsigma>2$\\
$M_{AA}(n_h,n_\varsigma)\sim M\left|\frac{\tilde {\mathcal K}_{AA}^{
(n_\varsigma)}}{\tilde {\mathcal K}_{AA}
^{(0)}}\right|^{\frac{-1}{n_h+n_\varsigma}},$
&$n_h\geq0, \;n_\varsigma>0$\\
$M_\varepsilon(n_h,n_\varsigma)\sim M\left|
\frac{\tilde {\mathcal K}_\varepsilon^{(n_\varsigma)}}{\tilde {\mathcal K}_{AA}
^{(0)}}\right|^{\frac{-1}{n_h+n_\varsigma}},$
&$n_h\geq0, \;n_\varsigma>0$\\{}
 \end{tabular}\\\hline
\end{tabular}\end{center}
\label{table2}
\caption{Relevant parameters appearing in Table 1}
\end{table}

It is clear that it would be desirable to have an independent argument  for
choosing the arbitrary TDF (e.g. an additional symmetry) such that they automatically satisfy the conditions in Table \ref{table}. 
For the moment,  we have unfortunately not
found such a rationale.
Nevertheless, we will give in this section three explicit {\it ad hoc} examples
to show the existence of TDF satisfying the previous requirements.

%%%%%%%%%%%%%%%%%%%%%%
\subsection{Polynomial TDF}\label{sec:poli}
%%%%%%%%%%%%%%%%%%%%%%
The first example we give is motivated by its simplicity. All
theory-defining functions can be taken to be polynomials of the 
metric determinant. In analogy with the Higgs potential we choose
\begin{equation}
\label{scpot}
v(-g)=\frac{\l}{4}\left(g_0^2-(-g)^2\right)^2\;,
\end{equation}
which satisfies condition C1. The simplest  choice for the
 remaining functions is given by
\begin{equation}\label{sc}\begin{array}{lclclcl}
f(-g)&=&{\mathcal G}_{gg}(-g)&=&{\mathcal G}_{AA}(-g)&=&1\;,\\
{\mathcal G}_{g\phi}(-g)&=&{\mathcal G}_{na}(-g)&=
&{\mathcal G}_\epsilon(-g)&=&0\;,\\
{\mathcal G}_{\phi\phi}(-g)&=&(-g)^2\;.
\end{array}
\end{equation}
For this choice of functions the parameters of the theory are
summarized in Table 3 ($\s_0=-g_0$).
\begin{table}[h]\begin{center}
\begin{tabular}[]{|c|l|c|}\hline
i.& Signs of kinetic terms&
 $
  \epsilon_\varsigma=\epsilon_\phi=\epsilon_A=1\;.
$\\\hline
ii.& Masses and relevant couplings &
 \begin{tabular}{lll}
$m_\varsigma^2=2\l \s_0^2M^2$
&$\k_\varsigma= 6\l \s_0 M$
&$\l_\varsigma =6\l$
\\
$m_A^2=e^2 \s_0^2 M^2$
&$\k_A=2e^2 \s_0 M$
&$\l_A=2e^2$
 \end{tabular}\\\hline
iii.&Suppression scales&
 \begin{tabular}{l}\vspace{-.4cm}\\
 $
M_{\phi\phi}(n_h,1)
\sim M\left(\frac{6+\s_0^2}{2\s_0}\right)^{\frac{1}{1+n_h}}\;,$\\
$ M_{\phi\phi}(n_h,2)
\sim M\left(\frac{6+\s_0^2}{2}\right)^{\frac{1}{2+n_h}}.
$
\end{tabular}\\\hline
\end{tabular}\end{center}
\caption{Parameters for TDF for Section 7.1.
% given in (\ref{scpot}) and (\ref{sc})
}
\end{table}

The conditions C1-C3 in Table \ref{table} are immediately satisfied by
this choice of TDF. The conditions C4-C7 are satisfied provided that
$0<\sigma_0\ll 1$ and that $0<e^2\lesssim 1/2$ and $0<\lambda \lesssim
1/6$. Finally, the condition C8 always holds, independently of the
parameter values. The small value of $\s_0$ is responsible for the
hierarchy between the Planck scale $M$ and the scales related to the
scalar and vector sectors. It is also interesting to observe that the
higher dimensional operators are suppressed below the Planck scale
independently of the value of $\s_0$.

We conclude that the theory given by the Lagrangian \eqref{LU1} with
TDF \eqref{scpot} and \eqref{sc} is almost equivalent to the
renormalizable Abelian Higgs model at energies well below the Planck
scale $M$. The only difference is the term coming from the dilaton, 
\be
\label{AH_dilaton}
{\mathcal L}_d=-\frac{1}{2}\sqrt{-\tilde g}((\tilde \s+\s_0)^2+6)(\pd \tilde \phi)^2.
\ee
The (non-renormalizable) interactions appearing in this term
certainly produce differences between  the two theories, but these effects
are suppressed both by the Planck scale and by the
derivative coupling of the dilaton. They may be relevant in the context of cosmology, discussed in the next section.

Finally, we would like to note that by changing variables one can easily find other sets
of polynomial functions which describe a theory equivalent to the one
given by \eqref{scpot} and \eqref{sc} (and thus also to the Abelian Higgs model) and which also satisfy all
conditions C1-C8. For example, one can redefine the metric and the
scalar field $\phi$ through\footnote{A slightly more general family equivalent
to the Abelian Higgs models in the previous sense is easily found by allowing a generic function of $\tilde\s$ in 
\eqref{AH_dilaton}.}
\begin{align}
g_{\m\n}&\mapsto\left(- g\right)^{2\a} g_{\m\n}\;,\\
\phi&\mapsto\left(- g\right)^{\b} \phi\;,
\end{align}
where $\a$ and $\b$ are some arbitrary numbers. In terms of the new
variables the Lagrangian \eqref{LU1} keeps its structure. The TDF
equivalent to \eqref{scpot} and \eqref{sc} are
\begin{equation}\label{scfamily}\begin{array}{lcl}
v(-g)&=&\dfrac{\l}{4}\left( g_0^{2+16\a}-\left(- g\right)^{2+16\a}\right)^2
\left(- g\right)^{4(\a+\b)}\;,\\
f(-g)&=&(- g)^{2(\a+\b)}\;,\\
{\mathcal G}_{gg}(- g)&=&\left((1+8\a)^2+\b^2\right)(- g)^{18\a+2\b}
-(6\a^2+12\a\b)(- g)^{2(\a+\b)-2}\;,\\
{\mathcal G}_{\phi\phi}(- g)&=&(- g)^{18\a+2\b+2}\;,\\
{\mathcal G}_{g\phi}(- g)&=&6\a(- g)^{2(\a+\b)-1}+\b(- g)^{18\a+2\b+1}\;,\\
{\mathcal G}_{AA}(- g)&=&1\;,\\
{\mathcal G}_{na}(- g)&=&{\mathcal G}_\epsilon(- g)=0\;.
\end{array}
\end{equation}
It is straightforward to check explicitly that for $0<(-g_0)^{1+8\a}\ll 1$,
$0<e^2\lesssim 1/2$ and $0<\lambda \lesssim
1/6$ this set of polynomials also satisfies the
conditions C1-C8. The two-parameter family of sets of functions
\eqref{scfamily} describes one and the same theory for different
variable choices. For $\a=\b=0$ the functions take the simple forms
\eqref{scpot} and \eqref{sc}.

%%%%%%%%%%%%%%%%%%%%%%%%%%%%%%%%%%%
\subsection{TDF leading to Abelian Higgs model plus a decoupled dilaton}
%%%%%%%%%%%%%%%%%%%%%%%%%%%%%%%%%%%
In this example we show that one can choose TDF such that the particle physics part of the theory is exactly the Abelian Higgs model and the dilaton only couples to the gravitational field.
To this end, we turn our attention to the Lagrangian in the form \eqref{Egauge} and notice that if the TDF are such that
\begin{equation}\label{AHreq}\begin{array}{lcl}
V(\s)&=&\dfrac{\l}{4}\left(\s^2-\s_0^2\right)^2\,,\\
f(\s)&=&\s^{-1/4}\,,\\
\mathcal{K}_{\s\s}(\s)&=&\mathcal{K}_{\phi\phi}(\s)=\mathcal{K}_{AA}(\s)=1\,,\\
\mathcal{K}_{\s\phi}(\s)&=&\mathcal{K}_{\s A}(\s)=\mathcal{K}_{\epsilon}(\s)=0\,,\\
\mathcal{K}_{int}(\s)&=&\s^2\,,
\end{array}\end{equation}
that Lagrangian reads
\begin{equation}\label{EgaugeAH}
\begin{split}
 \frac{\mathcal{L}_e}{\sqrt{-\tilde g}}=&-\frac{1}{2}M^2
\tilde R-\frac{1}{2}M^2(\partial\s)^2-\frac{1}{2}M^2(\partial\ln(\phi/M))^2
-\frac{1}{2}e^2M^2\s^2
A^2-\frac{1}{4}F^2
\\&-M^4\frac{\l}{4}\left(\s^2-\s_0^2\right)^2-\frac{M^4\Lambda_0}{\phi^4}\;.
\end{split}
\end{equation}
For this particular case, the transformations \eqref{trgauge} reduce to
$\tilde\s=\s$, $\tilde\phi=M\ln\frac{\phi}{M}$ and $\tilde A_\m=A_\m$ and \eqref{eq:Lagrgauge} becomes
\begin{equation}\label{EgaugeAHfin}
\begin{split}
 \frac{\mathcal{L}_e}{\sqrt{-\tilde g}}=&-\frac{1}{2}M^2
\tilde R-\frac{1}{2}M^2(\partial\tilde\s)^2-\frac{1}{2}(\partial\tilde\phi)^2
-\frac{1}{2}e^2M^2\tilde\s^2
\tilde A^2-\frac{1}{4}\tilde F^2
\\&-M^4\frac{\l}{4}\left(\tilde\s^2-\tilde\s_0^2\right)^2-\Lambda_0\exp\left(-\frac{4\tilde\phi}{M^4}\right)\;.
\end{split}
\end{equation}
This is the Lagrangian of the Abelian Higgs model, $\tilde\s$ being the Higgs field in the unitary gauge, plus a dilaton field $\tilde\phi$ with an exponential potential proportional to $\L_0$ and coupling only to gravity.

Making use of the relations \eqref{Ks} it is straightforward to find a set of TDF that satisfy the requirements \eqref{AHreq}:\footnote{Note that just
like in the above example, this set of functions is only one
representative of an infinite family of sets of functions that
correspond to the same theory.}
\begin{equation}\label{ahc}\begin{array}{lcl}
v(-g)&=&\dfrac{\l}{4}(-g)^{-1/2}(g_0^2-(-g)^2)^2\;,\\
f(-g)&=&(- g)^{-1/4}\;,\\
{\mathcal G}_{gg}(- g)&=&(- g)^{-1/4}-\dfrac{3}{32}(- g)^{-9/4}\;,\\
{\mathcal G}_{\phi\phi}(- g)&=&(- g)^{7/4}\;,\\
{\mathcal G}_{g\phi}(- g)&=&\dfrac{3}{4}(- g)^{-5/4}\;,\\
{\mathcal G}_{na}(- g)&=&-5(- g)^{-1/4}-(- g)^{7/4}\;,\\
{\mathcal G}_{AA}(- g)&=&1\;,\\
{\mathcal G}_{\epsilon}(- g)&=&0\;.
\end{array}
\end{equation}
By construction, this set of TDF satisfies all conditions C1-C8, as soon as $0<-g_0\ll 1$,
$0<e^2\lesssim 1/2$ and $0<\lambda \lesssim 1/6$. Moreover, as the theory corresponds to the Abelian Higgs model, its particle physics part contains no higher dimensional operators and is renormalizable.

The choice of TDF given by \eqref{ahc} might seem somewhat peculiar. However, one should remember that the explicit expressions of the TDF depend on the variables in which one chooses to express the Lagrangian. In particular, if one chooses variables such that $d_g=2$ and $d_\phi=0$ (c.f. Section \ref{sitd}), the arbitrary functions only depend on $\phi$. In terms of those variables, the Abelian Higgs model plus decoupled dilaton corresponds to choosing the arbitrary functions to be polynomials in $\phi$.

%%%%%%%%%%%%%%%%%%%%%%%%%%%%%%%%%%%
\subsection{TDF reproducing scale-invariant unimodular gravity}
%%%%%%%%%%%%%%%%%%%%%%%%%%%%%%%%%%%

In an earlier work \cite{Shaposhnikov:2008xb} two of us (M. S. and D.
Z.) presented a model which combines scale invariance and unimodular
gravity. There, a new singlet scalar field was introduced in order to
make both the gravitational and the matter part of the action scale-invariant. Unlike in the present proposal, that scalar field was
introduced \emph{ad hoc} and was not related to the restriction of the gauge group from Diff to TDiff. Due to the shape of the
potential, scale invariance was spontaneously broken. In the same
model, standard GR was replaced by unimodular gravity with the aim of
introducing a cosmological constant without explicitly breaking scale invariance.  As already
mentioned in Sections \ref{intro} and \ref{tdit}, the unimodular theory
can be considered as a particular TDiff model with the constraint $g=-1$. Therefore, the model of
\cite{Shaposhnikov:2008xb} can certainly be written as a scale-invariant TDiff theory.
To find the corresponding TDF, we will consider the simpler example where the full SM considered in \cite{Shaposhnikov:2008xb} is
replaced by
the Abelian Higgs model, analyzed in the present paper. After choosing the TDF as\footnote{Like in the above examples, this set of functions is only one
representative of an infinite family of sets of functions that
correspond to the same theory.}
\renewcommand\arraystretch{1.5}
\begin{equation}\label{oldmodel}\begin{array}{lcl}
v(-g)&=&\dfrac{\l}{4} \left(2-\zeta^2 (-g)^{-2}\right)^2\;,\\
f(-g)&=&\xi_\chi (-g)^{-2}+2\xi_h\;,\\
{\mathcal G}_{gg}(-g)&=&\dfrac{49-90\xi_\chi}{64}(-g)^{-4}
+\dfrac{1+6\xi_h}{32}(-g)^{-2}\;,\\
{\mathcal G}_{\phi\phi}(-g)&=&2\;,\\
{\mathcal G}_{g\phi}(-g)&=&-\dfrac{7-6\xi_\chi}{8}(-g)^{-3}
+\dfrac{1+6\xi_h}{4}(-g)^{-1}\;,\\
{\mathcal G}_{na}(-g)&=&(-g)^{-2}\;,\\
{\mathcal G}_{AA}(-g)&=&1\;,\\
{\mathcal G}_\epsilon(-g)&=&0\;,
\end{array}
\end{equation}
the Lagrangian \eqref{LU1} can be brought to
the form
\begin{equation}\label{old}
\begin{split}
\mathcal{L}_{SZ}=-\frac{1}{2}(\xi_\chi \chi^2+&2\xi_h\Phi\Phi^*)\hat R-\frac{1}{2}
\hat g^{\m\n}\partial_\m \chi\partial_\n \chi
-\hat g^{\m\n}D_\m\Phi(D_\n\Phi)^*\\
&-\frac{1}{4}\hat g^{\m\n}\hat g^{\r\s}F_{\m\r}
F_{\n\s}-\frac{\l}{4}(2\Phi \Phi^*-\zeta^2\chi^2)^2\;,
\end{split}
\end{equation}
where we have defined the unimodular metric $\hat
g_{\m\n}=(-g)^{-1/4}g_{\m\n}$ and the scalar fields
${\Phi=\phi(-g)^{1/8}}$ and ${\chi=|\phi |(-g)^{-7/8}}$. $\hat R$ is
the Ricci scalar associated to the unimodular metric $\hat g_{\m\n}$.
Note that for the variable change ${\chi=|\phi|(-g)^{-7/8}}$ to be well defined, $\chi$ is only allowed to take positive values. However, the theory being symmetric under
$\chi\mapsto-\chi$, one can equally allow for negative values of
$\chi$. In that part of phase space the matching of the variables is
${\chi=-|\phi|(-g)^{-7/8}}$. We see that \eqref{old} is exactly the Lagrangian of the model proposed in 
\cite{Shaposhnikov:2008xb} reduced to the Abelian Higgs model.

As for the choice of TDF discussed in the previous subsection, the choice of functions
\eqref{oldmodel} is rather peculiar and in particular, the
presence of the non-analytic term $\mathcal
G_{na}\neq0$ in  \eqref{LU1} is essential to find the Lagrangian
${\mathcal L}_{SZ}$. Again, there exists a set of variables, $\hat g_{\m\n}$, $\phi$ and $\chi$, in terms of which the expression of the Lagrangian becomes particularly simple.

The complex scalar field $\Phi$ in (\ref{old}) plays the role of the Higgs field,
non-minimally coupled to gravity. If one includes fermions, then this
is the field that couples to fermions through Yukawa couplings. The
real scalar field $\chi$ is a kind of dilaton. The flat direction in
the potential guarantees that the theory possesses an infinite family
of ground states which spontaneously break the dilatational symmetry.
In \cite{Shaposhnikov:2008xb} it was shown that the Lagrangian
\eqref{old} (if one adds all SM matter and gauge fields) represents
a viable model for SM phenomenology which besides
enforces
interesting cosmological phenomena if the parameters
are positive and such that $\zeta\lll 1$, $\xi_\chi\sim\mathcal{O}(10^{-3})$, $\xi_h\sim \mathcal{O}(10^{5})$
 and $\l\lesssim 1$. The smallness of $\zeta$ is responsible for the
hierarchy between the Planck scale and the electroweak scale. The
values of $\xi_\chi$ and $\xi_h$ are fixed by cosmological
considerations (cf. \cite{Shaposhnikov:2008xb}) .

Let us now check, whether the model given by \eqref{oldmodel} satisfies
the conditions C1-C8 appearing in Table \ref{table}. To this end, we
consider the expansion of the different  functions expanded around a
constant solution $g_0=\s_0=\frac{\sqrt{2}}{\zeta}$.  The different
parameters are summarized in Table \ref{siug}.%
\begin{table}\begin{center}
\begin{tabular}{|c|l|c|}
\hline
i.&  Signs of kinetic terms
 &$
  \epsilon_\varsigma=\epsilon_\phi=\epsilon_A=1\;.
$\\\hline
ii.& Masses and relevant couplings&
 \begin{tabular}{ll}
&$m_\varsigma^2=2\l\frac{\zeta^2}{\xi_\chi}M^2(1+\mathcal{O}(\zeta^2))$,\\
&$\k_\varsigma= 6\l \sqrt{\frac{\zeta^2}{\xi_\chi}}M(1+\mathcal{O}(\zeta^2))$,\\
&$\l_\varsigma =6\l(1+\mathcal{O}(\zeta^2))$,\\
&$m_A^2=e^2\frac{\zeta^2}{\xi_\chi}M^2(1+\mathcal{O}(\zeta^2))$,\\
&$\k_A=2e^2\sqrt{\frac{\zeta^2}{\xi_\chi}} M(1+\mathcal{O}(\zeta^2))$,\\
&$\l_A=2e^2(1+\mathcal{O}(\zeta^2))$.
 \end{tabular}
\\\hline
iii.& Suppression scales&
$
M_{\phi\phi}, M_V, M_{int}, M_{AA} \sim \frac{M}{\xi_h}<M\;.
$
\\\hline
\end{tabular}\end{center}
\caption{Parameters  the scale-invariant unimodular gravity (Section 7.2).}
\label{siug}
\end{table}
For the phenomenologically interesting parameters, conditions C1-C6 and C8 hold.
We are left with the question about higher dimensional operators. In the present example, all terms of \eqref{eq:Lagrgauge}, except the one
proportional to $\tilde {\mathcal K}_\epsilon(\tilde \s)=0$ and the
one proportional to $\Lambda_0=0$, give rise to an infinite number of higher
dimensional operators. Depending on the values of the parameters,
their suppression scales can be smaller than the Planck scale $M$. For
the phenomenologically interesting parameters,
the lowest suppression scales are of the order $\frac{M}{\xi_h}$.
Although significantly smaller than the Planck scale, this scale is
still much higher then the scales relevant to particle
physics and can be consider as the cut-off scale of the theory\footnote{Note that this lowering of the physical cut-off
scale below the Planck scale happens is generic in models where the
Higgs field is non-minimally coupled to gravity. For a recent
discussion on this issue see \cite{Bezrukov:2010jz}.}.
Although condition C7 is not exactly satisfied,
the higher dimensional operators are still negligible at particle
physics scales and the rest of conditions are fulfilled to high accuracy. We conclude that at energies well below
$\frac{M}{\xi_h}$ the theory given by the Lagrangian \eqref{LU1} with
defining functions \eqref{oldmodel} (respectively the equivalent
theory \eqref{old}) is also indistinguishable from the renormalizable
Abelian Higgs model.

%%%%%%%%%%%%%%%%%%%%%%%%%%%%
\section{The case $\Lambda_0\neq 0$, cosmology and dilaton interactions}
\label{cosmo}
%%%%%%%%%%%%%%%%%%%%%%%%%%%%

So far we have mainly considered static backgrounds for which
(\ref{maincond}) is satisfied and  $\Lambda_0=0$. The first
condition is about the TDF. It is equivalent to the absence of a
cosmological constant (cf. Section \ref{ldof}). The second condition is about the choice of the
initial state. It is related to the TDiff invariance and has nothing to do
with the TDF. Some motivations for the first condition in  (\ref{maincond}),
valid when gravity is dropped of (i.e. neglecting the scalar
curvature term $R$ in the Jordan frame action) were given in
\cite{Shaposhnikov:2008xi}. Namely, if $v(-g_0)>0$, the ground state of
the system is scale-invariant, meaning that the theory does not have
any particle excitations or that the theory is free. If  $v(-g_0)<0$,
the theory does not have a ground state at all. In other words, the
only sensible case is $v(-g_0)=0$ corresponding to a flat direction in the scalar potential
and leading to spontaneous breakdown of scale invariance.
As we have seen in Section \ref{sitd}, if gravity is included, the
cases where $v(-g_0) \neq 0$ do not have  known pathologies and simply
correspond to dS or AdS spaces, characterized by
a non-zero cosmological constant. So, scale-invariant TDiff theories do not give a
solution to the cosmological constant problem. Still, they provide
another perspective towards its solution, transferring the problem to
the requirement of some specific property (eq. (\ref{maincond})) of
one of the TDF.

We will start this section by considering that (\ref{maincond}) is satisfied,
but  $\Lambda_0 \neq 0$. Let us discuss qualitatively the
cosmological solutions in our theories and see how they affect local
particle physics. To this end we consider the Lagrangian
\eqref{fulltdscale_diff_ein} to which we add a matter part
\be
\begin{split}
\frac{\mathcal{L}_e}{\sqrt{-\tilde{g}}}=&
-\frac{1}{2}M^2\tilde{R}-\frac{1}{2}\e_\s M^2
(\partial\tilde\s)^2
-\frac{1}{2}\tilde{\mathcal K}_{\phi\phi}(\tilde{\s})
(\partial\tilde\phi)^2
\\
&-M^4\tilde{V}(\tilde{\s})-\Lambda_0 \tilde{\mathcal K}_{\Lambda_0}(\tilde
\s)\exp\left(-\frac{4\tilde\phi}{M}\right)+\mathcal{L}_m\;,
\label {Lcosmo}
\end{split}
\ee
where $\mathcal{L}_m$ contains all bosonic and fermionic degrees of freedom
of the SM coupled to the scalar fields and gravity in the way described
in Sections \ref{ctb} and \ref{ctm}. Notice that the dependence of the potential on
$\tilde\phi$ is \emph{uniquely} determined by the way
scale invariance is broken in TDiff theories. Similar potentials have
been considered in the past in the context of scalar tensor theory,
cf. \cite{Fujii:2003pa}. Consider now the homogeneous fields $\tilde
\sigma=\tilde \sigma(t)$ and $\tilde \phi=\tilde \phi(t)$ living in a
flat Friedmann-Lema\^itre-Robertson-Walker (FLRW) space-time with metric
\be
\di \tilde s^2=-\di t^2+\tilde a(t)^2 \di \vec{x}^2\;,
\ee
where $\tilde a(t)$ is the scale factor. The dynamics of the
homogeneous scalar fields is mainly determined by the potential
\be
\tilde V_{\Lambda_0}(\tilde \sigma,\tilde
\phi)=M^4 \tilde{V}(\tilde{\s})+\Lambda_0 
\tilde{\mathcal K}_{\Lambda_0}(\tilde \s)\exp\left(-\frac{4\tilde\phi}{M}\right)\;.
\ee
As long as the kinetic term of the scalar fields is positive-definite,
the scalar fields tend to roll down the potential, with some friction
caused by the expansion of space-time. In the $\tilde \sigma$-direction
the potential has a minimum at $\tilde \sigma=0$ due to
the conditions \eqref{maincond1}\footnote{Note that to
get this minimum it is enough that condition (\ref{minim_cond}) holds.}. In the $\tilde \phi$-direction, the
potential is governed by the exponential factor. If  $\Lambda_0 
\tilde{\mathcal K}_{\Lambda_0}(\tilde \s)>0$, the potential is of the
run-away type, i.e. it gets minimal for $\tilde
\phi\rightarrow\infty$. In the opposite -- pathological -- case the
potential for $\tilde\phi$ is not bounded from below.  Hence, a
typical evolution of the scalar condensates $\tilde \sigma$ and
$\tilde \phi$ will be the following: The first term of the potential
$\tilde V_{\Lambda_0}$ drives the trajectories towards the ``valley'' $\tilde
\sigma=0$. Due to the Hubble friction the field
undergoes damped oscillations around the valley before asymptotically
approaching $\tilde \sigma=0$. The second term in $\tilde V_{\Lambda_0} $
drives the trajectory towards $\tilde \phi\rightarrow\infty$. After
$\tilde\sigma$ has settled down in the valley, this leads to
a roll-down along the valley\footnote{We neglect here effects of
potential-terms involving couplings of  $\tilde\sigma$ to the
SM-fields. Also, we assumed that the function $\tilde{\mathcal
K}_{\Lambda_0}$ defined in (\ref{eq:tildepot})  does not play a
significant role in the cosmological evolution.}.

For appropriate choices of the TDF and initial
conditions, the roll-down \emph{towards} the valley $\tilde \sigma=0$
can give a mechanism for inflation. During the subsequence
roll-down \emph{along} the
valley, the scalar fields can play the role of a dynamical dark-energy component
(quintessence). This is a generic scenario for scale-invariant TDiff theories. 
A concrete realization has been
proposed in \cite{Shaposhnikov:2008xb} (see also \cite{inprep}).

Since the evolution drives $\tilde \sigma\rightarrow 0$ it
seems reasonable to assume that in the present universe $\tilde
\sigma\simeq 0$. If this is fulfilled, then all masses
and couplings of the SM-particles are like in the case $\Lambda_0=0$
described in the above sections. The only effects of the cosmological
background on particle physics would then come through
$\tilde\phi(t)$. One can put simple and still very strong bounds on
the influence of $\tilde\phi(t)$ by requiring that its energy density
does not give a too big contribution to the energy density of the
universe. In other words, both the kinetic and the potential energy of
the condensate $\tilde\phi(t)$ have to be smaller than today's critical
energy density $\r_{cr}^0=3M^2 H_0^2\simeq 10^{-120}M^4$, i.e.
\begin{align}
&\frac{1}{2}\tilde
{\mathcal K}_{\phi\phi}\left(\tilde\sigma_0=0\right)(\partial_0\tilde
\phi)^2<\r_{cr}^0,\\
&\Lambda_0 \tilde V(\tilde
\s_0=0)\exp\left(-\frac{4\tilde\phi}{M}\right)<\r_{cr}^0\;.
\end{align}
These constraints, together with the conditions on the derivatives of
$\tilde {\mathcal K}_{\phi\phi}(\tilde \s)$,  \eqref{supcond1} and
similar conditions on the derivatives of $\tilde V_{\Lambda_0}(\tilde \s)$
guarantee that all interactions induced by ${\partial_0\tilde \phi\neq
0}$ are highly suppressed and can be neglected in the description of
local particle interactions.

Finally,  we would like to briefly comment on the situation where $v(\s_0)\neq 0$.
As we already stated, this case is equivalent to the presence of a cosmological
constant (cf. (\ref{vacu_scale})). In this situation, phenomenological bounds 
imply that this term must be very small, and will not affect
the conclusions on local physics of the previous sections. It will, however, 
be important for late time cosmology, as it represents a contribution to dark energy on top
of those coming from the dynamics of the scalar fields presented previously in this section.
In fact, asymptotically, this constant term becomes dominant over the other contributions, as they
are diluted during the expansion of the universe.

%%%%%%%%%%%%%%%%%%%%%%%%%
\section{Conclusions}
\label{concl}
%%%%%%%%%%%%%%%%%%%%%%%%%

In this paper we have shown that scale-invariant TDiff theories constitute
a viable alternative to standard General Relativity (GR). The group of space-time symmetries
of these theories is not the full group of diffeomorphisms, but rather its subgroup
defined by 4-volume preserving transformations. Hence, TDiff theories depend on a number of a priori arbitrary functions of the metric determinant, the theory-defining functions (TDF). As a consequence, TDiff theories generically have more physical degrees of freedom in the gravitational sector: in addition to the massless graviton they contain a propagating scalar degree of freedom that may or may not be massive. 

In order to study the phenomenology of TDiff theories, we first formulated them in terms of equivalent Diff invariant theories by means of a St\"uckelberg field.
An advantage of the St\"uckelberg formalism is that it makes the new scalar field appear explicitly in the Lagrangian. A very interesting feature of TDiff theories is the appearance of an arbitrary mass scale
$\L_0$.  In the TDiff formulation of the theory this scale appears as an integration constant in the equations of motion, while in the equivalent Diff invariant formulation it appears as a new coupling constant in the Lagrangian. The appearance of $\L_0$ is exactly analog to the appearance of an arbitrary cosmological constant in unimodular gravity. Notice, however, that in the present context $\L_0$ does not play the role of a cosmological constant. 

Next, we focused on the scale-invariant case, i.e. we considered actions that are invariant under global dilatations. We were interested in the situation where scale invariance is spontaneously broken, such that all scales of the theory are induced by the expectation value of a scalar field. We found that if the theory contains only one scalar degree of freedom, this degree of freedom is necessarily the Goldstone boson of the spontaneously-broken scale invariance. Therefore, as our objective was to construct a theory in which a scalar field plays the role of the SM Higgs field, we were lead to the introduction of an additional scalar field. The scale-invariant TDiff theories including an additional real scalar field were studied in detail.
After extending the scale-invariant TDiff theories to gauge theories and including fermionic fields, we discussed how the framework can be generalized to include all degrees of freedom of the Standard Model.

For $\L_0=0$, the spectrum of scalar excitations around a symmetry-breaking background consists of the  a massless scalar dilaton plus a potentially massive scalar degree of freedom. 
The dilaton decouples from all SM degrees of freedom except for the Higgs field, to which it couples derivatively.
A non-zero $\Lambda_0$ leads to a very particular potential term breaking the scale symmetry explicitly.
This term can yield an additional interaction between the dilaton and the Higgs field which is, however, negligible for particle physics phenomenology. Interestingly, the $\L_0$-term can depend on the dilaton only through the exponential function $\exp(-4\tilde\phi/M)$. As a consequence, the dilaton can give rise to dynamical dark energy.

For the theory restricted to the gravitational, vector and scalar sectors, we derived the conditions on the TDF leading to a renormalizable low-energy theory in the particle physics sector. Moreover, we gave three explicit examples of TDF satisfying these conditions. One of the examples corresponds to the model of \cite{Shaposhnikov:2008xb}.

Next, we commented on the generic behavior of cosmological solutions. In particular, we found that the conditions yielding a theory close to the SM, entail an interesting cosmological phenomenology. Namely, the corresponding solutions can describe a phase of inflation in the very early universe, whereas the existence of a small $\L_0$ produces a run-away potential for the dilaton, 
which can hence play the role of a dynamical dark energy component.

Finally, generic TDF also imply the presence of a pure cosmological constant term. Keeping this contribution small represents a fine-tuning
similar to the fine-tuning required in GR to get a small cosmological constant. 
Indeed, it is interesting to compare this situation with the naturalness problem of the Higgs mass. 
Scale invariance is a key ingredient for a solution of the problem
of stability of the Higgs mass against radiative corrections. This
invariance keeps the corrections small (at least if dimensional regularization is used \cite{Shaposhnikov:2008xi}). At the same time, the smallness of the Higgs mass in
comparison with the Planck scale is not explained and must be imposed
``by hand'', as in SM with a cut-off given by the scale of strong coupling of GR. The same statement is true for the
cosmological constant. In spite of the fact that scale invariance forbids any mass
parameters in a theory, a cosmological constant is generally present, and is
related to the self-interaction of a scalar field (Higgs) in the J-frame.
Only tuning this term to zero makes for the absence of a cosmological constant.

\section*{Acknowledgements} This work was supported by the Swiss
National Science Foundation and by Tomalla Foundation. We thank Jaume
Garriga and Sergey Sibiryakov 
for helpful discussions.

%%%%%%%%%%%%%%%%%%%%%%%%%%%

\end{document}